\documentclass[11pt]{article}
\pdfoutput=1
\usepackage{jheppub,slashed,enumitem} 

\DeclareFontFamily{U}{wncy}{}
\DeclareFontShape{U}{wncy}{m}{n}{<->wncyr10}{}
\DeclareSymbolFont{mcy}{U}{wncy}{m}{n}
\DeclareMathSymbol{\Sha}{\mathord}{mcy}{"58} 
\def\unit{{\hbox{\kern+.5mm 1\kern-1mm l}}} 

\title{\textbf{On Finite-Size D-Branes in Superstring Theory}} 

\abstract{We test exact marginality of the deformation describing the blow-up of a zero-size D(-1) brane bound to a background of D3-branes by analyzing the equations of motion of superstring field theory to third order in the size. In the process we review the derivation of the instanton profile from string theory, extending it to include $\alpha'$- corrections.
}

\author{Luca Mattiello and}
\author{Ivo Sachs}
\affiliation{Arnold Sommerfeld Center for Theoretical Physics,\\ Ludwig Maximilian University of Munich,\\
Theresienstr. 37, D-80333 M\"unchen, Germany}
\emailAdd{Luca.Mattiello@physik.uni-muenchen.de}
\emailAdd{Ivo.Sachs@physik.uni-muenchen.de}

\begin{document}
\begin{flushright}
LMU-ASC 10/19\\
\end{flushright}
\maketitle

\section{Introduction}
It is well known that four dimensional, perturbative Yang-Mills theory with maximal supersymmetry (SYM) has a worldsheet description in terms of the superconformal sigma model (SCFT) of type II superstring theory on a D3 brane. Furthermore, pointlike instantons have an equivalent  description as bound states of D(-1) branes localized inside this D3 brane \cite{Witten:1995im, Douglas:1996uz,Douglas:1995bn, Billo:2002hm}. Consequently, these configurations, in spite of being non-perturbative from the four-dimensional, or D3 brane, point of view, have a simple description in terms of a worldsheet SCFT with mixed Neumann and Dirichlet boundary conditions \cite{Polchinski:1995mt}. In particular, the SCFT on the D(-1) brane reproduces the ADHM constraints for zero size instantons \cite{Witten:1995gx, Dorey:2002ik, Atiyah:1978ri, Hashimoto:2005qh}. 

In Yang-Mills theory the size of the instanton is a modulus that is protected, at quantum level, thanks to supersymmetry (SUSY). A natural question is then whether there is a corresponding deformation of the worldsheet theory that describes D(-1) branes of finite size. To lowest, non-linear order in the deformation, the answer is affirmative: there is a marginal deformation of the SCFT obtained when certain massless fields in the Hilbert space of strings stretched between the D(-1) and the D3 branes take a non-vanishing expectation value.  Furthermore, evaluating the gluon vertex in the perturbed SCFT to second order in the shifted background reproduces the asymptotic form of the instanton profile in the singular gauge \cite{Billo:2002hm}\footnote{Strictly speaking, since this an off-shell problem, this calculation has to be performed in string field theory, but at leading order in the large distance asymptotic expansion the profile of the on-shell world-sheet prediction is correct.}. 

At next to leading order (third order in size), in addition to the usual subtleties of conformal perturbation theory (e.g. \cite{Recknagel:2013uja}), one encounters subtleties in the integration of odd moduli in supermoduli space (e.g. \cite{Witten:2012ga}). In the textbook treatment of scattering amplitudes in the worldsheet approach the integration over odd moduli is implemented by using a variety of pictures for the external states. While standard arguments imply that on-shell amplitudes computed in this way generally do not depend on how the picture is distributed (eg.  \cite{Polchinski_book}), this is not necessarily so when some internal states go on-shell and furthermore, one is not guaranteed that any choice of pictures gives the correct result. The problem at hand is precisely a case where this situation arises and where the usual worldsheet approach is incomplete. 

In order to decide whether blowing up of the D(-1) brane is an exact modulus of string theory we then refer to super string field theory (SFT). Luckily for the open string a consistent classical SFT exists \cite{Berkovits:1995ab,Erler:2013xta,Kunitomo:2015usa}. Any consistent SFT can be taken to address this problem; here we will work with the $A_\infty$-SFT \cite{Erler:2013xta}, since it is formulated on the small Hilbert space, but other choices are possible (see \cite{Maccaferri:2018vwo}). Concretely, we will analyze the string field theory equation of motion derived form the $A_\infty$-SFT action to third order in the perturbation to determine if the linearized solution, or marginal deformation, corresponding to the blow up mode can be integrated. We show that there is no obstruction at second order in the size $\frac{\rho}{\sqrt{\alpha'}}$ even without imposing the ADHM constraints, at this order. We then calculate the instanton profile to this order as a closed function in $\frac{x}{\alpha'}$. At very large distance form the location of the D(-1) it converges to the on-shell calculation of \cite{Billo:2002hm}. At third order in the size $\frac{\rho}{\sqrt{\alpha'}}$ we find that the size modulus is obstructed due to a contact term that arises from an integral over an odd modulus in supermoduli space. 
However, this obstruction can be removed by an appropriate zero-momentum gluon background as previously suggested in \cite{Maccaferri:2018vwo} in the case of Berkovits' superstring field theory.\footnote{See also \cite{Maccaferri:2019ogq,Jakub} for a recent derivation in the $A_\infty$ theory.} Consequently, the corrected instanton profile receives a constant contribution proportional to the size $\rho$ of the instanton.  

One may wonder how the situation compares to the bosonic string. In this case standard world-sheet approach is applicable but the modulus is obstructed as shown in \cite{Mattiello:2018kue} because the marginal operator that implements the blow up fails to be exactly marginal, at least if the compactification radius is a multiple of the self-dual radius. 

The rest of the paper is organized as follows. In section \ref{sec:SYM}, in order to be self-contained, we briefly review the maximally supersymmetric Yang-Mills theory in four euclidean dimensions, with a particular focus on instanton solutions, both in regular and singular gauge. In section \ref{sec:brane_system} we review the D(-1)-D3 brane SCFT with focus on the massless excitations and their field theory limit. In section \ref{sec:marginal_operators} we describe the marginal operators generating a blow up of the size of a D(-1) brane bound to D3 branes. In section \ref{sec:SFT} we first review some aspects of the $A_\infty$-SFT required to analyze exact marginality of the blow-up mode and then establish  marginality at second order. In section \ref{subs:profile} we then  compute the first order contribution to the amplitude for the emission of a gluon from the D(-1)-D3 worldsheet and relate it to the instanton profile in $\mathcal{N}=4$ SYM theory. In section \ref{sec:third_order} we discuss the marginality of the blow-up modulus at third order in the size by analyzing the SFT-equations of motion to this order.  Finally we present our conclusions.  Appendix \ref{app:conventions} is devoted to notations and conventions, while appendix \ref{app:OPE} treats the property of operators used in this paper. In appendix \ref{app:profile} the detailed calculation of the instanton profile is analyzed and in appendix \ref{Pder} the detailed derivation of the contact terms relevant for section \ref{sec:third_order} is outlined. In appendix \ref{app:anomaly} we discuss anomalous contributions due to the presence of non-primary operators.

\section{\texorpdfstring{$\mathcal{N}$}{N}=4 Super Yang-Mills Theory and Instantons}\label{sec:SYM}

In this section we briefly review the $\mathcal{N}=4$ SYM theory with special focus on instanton solutions. The action of this theory can be obtained, for example, through dimensional reduction of $\mathcal{N}$=1 SYM theory in ten dimensions, which is fairly simple. The resulting effective action is \cite{Detournay:2009mp, Vandoren:2008xg}
\begin{equation}\label{eq:Effective4}
\begin{split}
\mathcal{S}_{SYM}=&\dfrac{1}{g_{YM}^2}\int d^4x \;\text{Tr}\bigg\lbrace \dfrac{1}{2}F_{\mu\nu}^2-2\bar{\Lambda}_{\dot{\alpha}A}\bar{\slashed{D}}^{\dot{\alpha}\beta}\Lambda_\beta^{\;\;A}+(D_\mu\varphi_a)^2-\dfrac{1}{2}[\varphi_a,\varphi_b]^2\\
&-i(\Sigma^a)^{AB}\bar{\Lambda}_{\dot{\alpha}A}[\varphi_a,\bar{\Lambda}^{\dot{\alpha}}_{\;\;B}]-i(\bar{\Sigma}^a)_{AB}\Lambda^{\alpha A}[\varphi_a,\Lambda_{\alpha}^{\;\;B}] \bigg\rbrace\,.
\end{split}
\end{equation}
The indices $\mu,\nu=1,\dots,4$ are spacetime indices, while $\alpha,\dot{\alpha}=1,2$ are chiral and anti-chiral spinor indices. The index $a$ runs from 1 to 6, while the indices $A,B$ are spinor indices in 6 dimensions (corresponding to the six compactified directions). Details on the gamma matrices used to define $\slashed{D}$ can be found in appendix \ref{app:conventions}. $\Sigma^a$ and $\bar{\Sigma}^a$ are matrices that realize the six-dimensional Clifford algebra. The fields described by this action are a gauge field $A_\mu$, with field strength $F_{\mu\nu}=\partial_\mu A_\nu-\partial_\nu A_\mu+[A_\mu,A_\nu]$, 4 pairs of Weil spinors (gauginos) $\Lambda^{\alpha A}$ and $\bar{\Lambda}_{\dot{\alpha}A}$, respectively right- and left-handed and 6 scalars $\varphi_a$.

\subsection*{Instantons in $\mathcal{N}$=4 SYM theory}
When the gauginos and the scalars are not present the equations of motion simplify to
\begin{equation}
D^\nu F_{\nu\mu}=0\,,
\end{equation}
which are the same ones characterizing a pure Yang-Mills theory. In Euclidean signature, solutions of these equations are given by instanton, characterized by a self-dual or anti-self-dual field strength 
\begin{equation}
F_{\mu\nu}=\pm\dfrac{1}{2}\epsilon_{\mu\nu\rho\sigma}F^{\rho\sigma}=\pm \widetilde{F}_{\mu\nu}
\end{equation}
and by the winding number (or Pontryagin class) $k$.
In the following we focus on the simplest non-abelian gauge group, $SU(2)$. In the case of a higher rank gauge group, like $SU(N)$, instantons can be obtained by embedding the $SU(2)$ solutions. All the fields, in particular the gauge field, belong here to the adjoint representation of $SU(2)$, thus they carry an index $c=1,2,3$, i.e. 
\begin{equation}
A_\mu=A_\mu^c T^c\,,
\end{equation}
where $T^c$ are the generators of the $\mathfrak{su}(2)$ algebra; we will use also $T^c=\frac{\tau^c}{2i}$, where $\tau^c$ are the usual Pauli matrices. Instanton  solutions are abundantly considered in the literature (see for example \cite{Belavin:1975fg} or \cite{Vandoren:2008xg}), and we will simply give the explicit solutions (for winding number $k=1$) here. For $SU(2)$ the solution is usually given in two different gauges. In the \textit{regular gauge} we have
\begin{equation}\label{eq:regular}
\begin{split}
&A_\mu^c(x;x_0,\rho)=2\dfrac{\eta^c_{\mu\nu}(x-x_0)^\nu}{(x-x_0)^2+\rho^2}\,, \\
&A_\mu(x;x_0,\rho)=-\dfrac{\sigma_{\mu\nu}(x-x_0)^\nu}{(x-x_0)^2+\rho^2}\,.
\end{split}
\end{equation}
The solution depends on five parameters (moduli): the position $x_0^\mu$ and the size of the instanton $\rho$. Here, $\eta^c_{\mu\nu}$ are the 't Hooft symbols defined in Appendix \ref{app:conventions}. The corresponding field strength is given by
\begin{equation}
F^c_{\mu\nu}=-4\eta^c_{\mu\nu}\dfrac{\rho^2}{[(x-x_0)^2+\rho^2]^2}\,;
\end{equation}
from this expression one can immediately see that the field strength is self-dual and that the winding number is $k=1$. The anti-instanton solution can be found replacing $\eta^c_{\mu\nu}$ with $\bar{\eta}^c_{\mu\nu}$; in that case one has $k=-1$.
In the \textit{singular gauge} we have instead
\begin{equation}\label{eq:singular}
A_\mu^c(x;x_0,\rho)=2\bar{\eta}^c_{\mu\nu}\dfrac{\rho^2(x-x_0)^\nu}{(x-x_0)^2[(x-x_0)^2+\rho^2]}\,.
\end{equation}
Despite the presence of the anti-self-dual symbol $\bar{\eta}^c_{\mu\nu}$, this solution has a self-dual field strength and $k=1$. This expression is singular at the point $x_0$. Since the singular gauge is better suited for comparison with string theory, we will consider this gauge in what follows.

\section{The D3-D(-1) Brane System}\label{sec:brane_system}
In this section we review the string theory setup we will use in order to describe instantons in $\mathcal{N}=4$ SYM theory in 4 dimensions. Such a setup consists of a bound state of $N$ D3 branes and $k$ D(-1) branes \cite{Witten:1995im,Douglas:1996uz} (equivalent descriptions involve bound states of D$p$ and D$(p+4)$ branes). This configuration can describe instantons with winding number $k$ in a theory with gauge group $SU(N)$. The bosonic coordinates $X^M$ and $\psi^M$ ($M=0,\dots,9$) obey different boundary conditions depending on the type of boundary: on the D(-1) branes all the coordinates satisfy Dirichlet boundary conditions, while on the D3 branes the first coordinates $X^\mu$ and $\psi^\mu$ ($\mu=0,\dots,3$) satisfy Neumann conditions and the remaining $X^a$ and $\psi^a$ ($a=4,\dots,9$) satisfy Dirichlet conditions.

\subsection{Vertex Operators}
In the system we are considering, there are four types of open strings: those stretching between two D3 branes (3/3 strings), those stretching between two D(-1) branes ((-1)/(-1) strings) and finally those with one endpoint on a D3 brane and the other one on a D(-1) brane (3/(-1) and (-1)/3 strings). We have to consider each type of string separately, as each one has its own spectrum and properties. Let us consider first of all the 3/3 strings. In the following we focus only on Neveu-Schwarz (NS) states. The massless NS states can be divided in a four-vector $A^\mu$ and six scalars $\varphi^a$; the corresponding (unintegrated and with a $c$-ghost) vertex operators in the canonical picture are:
\begin{equation}\label{eq:vertex_33_boson}
\begin{split}
V_{A}(z;k)=A_\mu c(z)\psi^\mu(z)e^{-\phi(z)}e^{ik\cdot X(z)}\,,\\
V_{\varphi}(z;k)=\varphi_a c(z)\psi^a(z)e^{-\phi(z)}e^{ik\cdot X(z)}\,,
\end{split}
\end{equation}
where the momentum $k^\mu$ is ingoing and flows only along the D3 brane; the polarization satisfies the trasversality condition $A_\mu k^\mu=0$. These states  (together with their partners in the Ramond sector) reproduce exactly the fields of $\mathcal{N}$=4 SYM in four dimensions. If $N$ is greater than 1, these vertex operators must be multiplied by a $N\times N$ Chan-Paton factor $(T^c)^{uv}$, in order to take into account all the possible D3 branes on which the endpoints of the strings can lie. Here $c$ is a $SU(N)$ colour index; therefore all the polarizations will transform in the adjoint representation of $SU(N)$, as expected. In what follows we will assign Chan-Paton indices directly to the polarizations if needed (for example we will write $A_\mu^{uv}$).

Let us now consider (-1)/(-1) strings; the situation is different from the 3/3 case, because now there are no longitudinal Neumann direction. Therefore, the corresponding states do not carry momentum, and have to be considered as moduli rather than dynamical fields. Among the 10 scalars of the NS sector, it is convenient to divide the ones corresponding to the longitudinal directions of the D3 branes from the others; their vertex operators are:
\begin{equation}\label{eq:vertex_-1-1_boson}
\begin{split}
V_{a}(z)=a_\mu c(z)\psi^\mu(z)e^{-\phi(z)}\,,\\
V_{\chi}(z)=\chi_a c(z)\psi^a(z)e^{-\phi(z)}\,.
\end{split}
\end{equation}
Again, we have not written explicitly the indices labelling between which of the $k$ D(-1) branes the string is stretching; we should add to all the vertex operators a $k\times k$ Chan-Paton matrix $(t^U)^{ij}$ with indices $i,j=1,\dots,k$. Here $U$ is a $SU(k)$ colour index.

Finally we consider 3/(-1) and (-1)/3 strings. In this case the four directions $\mu=0,\dots,4$ are characterized by mixed boundary conditions. This means that the fields corresponding to these strings do not carry momentum; furthermore the fields $\psi^\mu$ have integer-moded expansion in the NS sector, and not in the R sector as for the 3/3 or (-1)/(-1) strings. A closer analysis of these strings (see \emph{e.g.} \cite{Hashimoto:1996he}) shows that in the NS sector one has two bosonic Weyl spinors of SO(4), $w$ and $\bar{w}$, with vertex operators given by
\begin{equation}\label{eq:vertex_3-1_NS}
\begin{split}
V_w(z)=w_{\dot{\alpha}}c(z)\Delta(z)S^{\dot{\alpha}}(z)e^{-\phi(z)}\,,\\
V_{\bar{w}}(z)=\bar{w}_{\dot{\alpha}}c(z)S^{\dot{\alpha}}(z)\bar{\Delta}(z)e^{-\phi(z)}\,.
\end{split}
\end{equation}
Notice that only the anti-chiral spin fields $S^{\dot{\alpha}}$ appear in these vertex operators, the chirality being fixed by the GSO projection. It is also possible to choose the opposite chirality; this would correspond to studying the anti-self-dual instantonic solutions, instead of the self-dual one. $\Delta$ and $\bar{\Delta}$ are the bosonic twist and anti-twist fields; they have conformal dimension 1/4, and they change the boundary conditions of the four $X^\mu$ coordinates. One can express the twist field $\Delta$ as product of four twist fields $\sigma$ corresponding to each direction longitudinal to the D3 branes:
\begin{equation}
\Delta(z)=\sigma^0(z)\sigma^1(z)\sigma^2(z)\sigma^3(z)\,.
\end{equation}
Other details on these fields can be found in the literature, for example in \cite{Hashimoto:1996he}, \cite{Zamolodchikov:1987ae} and \cite{Mattiello:2018kue}. We also have to add to all the vertex operators in \eqref{eq:vertex_3-1_NS} a matrix $\zeta^{ui}$ (or $\bar{\zeta}^{iu}$) with $N\times k$ (or $k\times N$) entries, corresponding to all the possible pairs of D3 and D(-1) branes.

In order to give to all these vertex operators their canonical dimension, one should multiply them by a suitable prefactor containing the right power of $\alpha'$. All NS states should have dimension (length)$^{-1}$, therefore one should consider prefactors proportional to $\sqrt{\alpha'}$ \cite{Billo:2002hm,DiVecchia:1996uq}.

\subsection{Tree-Level Amplitudes, Effective Actions and ADHM Constraints}
From the setup consisting of $N$ D3 branes and $k$ D(-1) branes it is possible to derive the corresponding effective action in the following way. One should first compute all string scattering amplitudes involving massless string states, using the vertex operators defined above. One should then find an effective Lagrangian able to reproduce these amplitudes. Since the string tension $\alpha'$ is the only dimensionful constant, an expansion of the action in number of derivatives corresponds to an expansion in powers of $\sqrt{\alpha'}$. The low-energy effective action is the one resulting from the field theory limit $\alpha'\rightarrow 0$.

When dealing with string scattering amplitudes in the presence of two different sets of D-branes, it is important to specify what kind of correlation function one is considering. For example, a scattering amplitude involving only 3/3 strings must be normalized with the disk amplitude with the boundary conditions of a D3 brane (see \cite{Erler:2014eqa, Mattiello:2018kue} for related  discussions). For example, the scattering amplitude of a gauge vector and two gauginos is given by (see \cite{Billo:2002hm})
\begin{equation}
\langle\langle V_{\bar{\Lambda}}V_{A}V_{\Lambda}\rangle\rangle_{D3}=C_{4}\langle V_{\bar{\Lambda}}V_{A}V_{\Lambda}\rangle\,,
\end{equation}
where $\langle V_{\bar{\Lambda}}V_{A}V_{\Lambda}\rangle$ is the pure CFT correlation function and $C_4=\langle\langle \unit\rangle\rangle_{D3}$. Correlation functions of (-1)/(-1) strings, on the other hand, must be normalized with the prefactor $C_0=\langle\langle \unit\rangle\rangle_{D(-1)}$. The values of $C_4$ and $C_0$ can be computed using unitarity methods, the results being (see \emph{e.g.} \cite{DiVecchia:1996uq})
\begin{equation}
C_4\propto \dfrac{1}{g_{YM}^2\alpha^{'2}}\,, \qquad C_0\propto\dfrac{1}{g_{YM}^2}\,,
\end{equation}
where $g_{YM}$ is the (adimensional) gauge coupling constant of the four-dimensional euclidean theory. In fact, the full low-energy effective field theory corresponding to the massless 3/3 interactions is
\begin{equation}
\begin{split}
\mathcal{S}_{SYM}=&\dfrac{1}{g_{YM}^2}\int d^4x \;\text{Tr}\bigg\lbrace \dfrac{1}{2}F_{\mu\nu}^2-2\bar{\Lambda}_{\dot{\alpha}A}\bar{\slashed{D}}^{\dot{\alpha}\beta}\Lambda_\beta^{\;\;A}+(D_\mu\varphi_a)^2-\dfrac{1}{2}[\varphi_a,\varphi_b]^2\\
&-i(\Sigma^a)^{AB}\bar{\Lambda}_{\dot{\alpha}A}[\varphi_a,\bar{\Lambda}^{\dot{\alpha}}_{\;\;B}]-i(\bar{\Sigma}^a)_{AB}\Lambda^{\alpha A}[\varphi_a,\Lambda_{\alpha}^{\;\;B}] \bigg\rbrace\,,
\end{split}
\end{equation}
which is exactly the action of the four-dimensional $\mathcal{N}=4$ SYM theory \eqref{eq:Effective4}, with $F_{\mu\nu}=\partial_\mu A_\nu-\partial_\nu A_\mu+[A_\mu,A_\nu]$. An alternative choice is to rescale  all the vertex operators by $g_{YM}$; the corresponding effective action would be the same as \eqref{eq:Effective4}, but without the prefactor $g_{YM}^{-2}$, and with $F_{\mu\nu}=\partial_\mu A_\nu-\partial_\nu A_\mu+g_{YM}[A_\mu,A_\nu]$. The same procedure can be repeated for the (-1)/(-1) strings and the mixed strings. Since the normalization of the corresponding amplitudes is $C_0$, the resulting effective action will be of the form
\begin{equation}\label{eq:Effective0}
\mathcal{S}_{\text{moduli}}=\dfrac{1}{g_{0}^2}\text{tr}\bigg\lbrace -\dfrac{1}{4}[a_\mu,a_\nu]^2+ \dots \bigg\rbrace\,,
\end{equation}
where we have written explicitly only one term, and the trace is over $SU(k)$ and not $SU(N)$ as before. We have highlighted one particular term of the action, in order to discuss the role of the prefactor $g_0^{-2}$, where $g_0$ is the coupling of a zero-dimensional $SYM$ theory. Unlike the gauge coupling $g_{YM}$, however, $g_0$ is a dimensionful constant, that can be expressed as $g_0\propto g_{YM}/\alpha'$. Therefore the field theory limit $\alpha'\rightarrow 0$ is problematic. The solution to this issue is to rescale some of the moduli with a prefactor $g_0$, as explained in \cite{Billo:2002hm}, in order to obtain a well defined low-energy action for the moduli. For the mixed string, for example, the vertex operators one should use are
\begin{equation}\label{eq:rescaling}
\begin{split}
&V_a\sim g_0\sqrt{\alpha'}a_\mu c\psi^\mu e^{-\phi}\sim \dfrac{g_{YM}}{\sqrt{\alpha'}}a_\mu c\psi^\mu e^{-\phi}\,,\\
&V_w\sim g_{0}\sqrt{\alpha'}w_{\dot{\alpha}}c\Delta S^{\dot{\alpha}}e^{-\phi}\sim \dfrac{g_{YM}}{\sqrt{\alpha'}}w_{\dot{\alpha}}c\Delta S^{\dot{\alpha}}e^{-\phi}\,,\\
&V_{\bar{w}}\sim g_{0}\sqrt{\alpha'}\bar{w}_{\dot{\alpha}}c S^{\dot{\alpha}}\bar{\Delta} e^{-\phi}\sim \dfrac{g_{YM}}{\sqrt{\alpha'}}\bar{w}_{\dot{\alpha}}c\bar{\Delta} S^{\dot{\alpha}}e^{-\phi}\,.
\end{split}
\end{equation}
Since the vertex operators for the moduli should be dimensionless, this means that the polarizations of the vertex operators are dimensionful. In this case $a$, $w$ and $\bar{w}$ have dimension (length)$^{1}$, and are associated to the position and size of the instanton. After this rescaling, the limit $\alpha'\rightarrow 0$ (with $g_{YM}$ held fixed) is well defined. This also means that we are considering the limit $\alpha'\rightarrow 0$ with the size of the instanton kept constant. The final result for the moduli action is given in \cite{Billo:2002hm}, in terms of some auxiliary fields. The equations of motion of these auxiliary fields give rise to some constraints on the moduli. In particular we have
\begin{equation}\label{eq:ADHM}
\bar{\eta}_c^{\mu\nu}\left([a_\mu,a_\nu]+\frac{1}{2}\bar{w}_{\dot{\alpha}}(\bar{\sigma}_{\mu\nu})^{\dot{\alpha}\dot{\beta}}w_{\dot{\beta}}\right)=0\,,\\
\end{equation}
which is the bosonic ADHM constraint \cite{Atiyah:1978ri}. Let us restrict to the case $N=2$ and $k=1$ for simplicity. Since $k=1$, $a_\mu$ are just numbers, therefore $[a_\mu,a_\nu]=0$ and the constraint becomes $\bar{w}_{\dot{\alpha}}(\bar{\sigma}_{\mu\nu})^{\dot{\alpha}\dot{\beta}}w_{\dot{\beta}}=0$. The matrix $(\bar{\sigma}_{\mu\nu})^{\dot{\alpha}\dot{\beta}}$ is symmetric, hence we can parametrize a generic solution as $\bar{w}_{\dot{\alpha}}w_{\dot{\beta}}=\rho^2\epsilon_{\dot{\alpha}\dot{\beta}}$, where $\rho$ has dimension (length)$^1$, and corresponds to the size of the instanton, as we will see later. For $SU(2)$ an explicit solution to the constraint is given by
\begin{equation}\label{eq:ADHM_sol}
\bar{w}_{\dot{1}}=(\rho,0)\,,\quad \bar{w}_{\dot{2}}=(0,\rho)\,,\quad w_{\dot{1}}=\begin{pmatrix}0\\-\rho\end{pmatrix}\,,\quad w_{\dot{2}}=\begin{pmatrix}\rho\\0\end{pmatrix}\,.
\end{equation}

\section{Marginal Vertex Operators}\label{sec:marginal_operators}
Out of all the vertex operators introduced in the previous section, we can identify some which are marginal, i.e. vertex operators of conformal dimension 0 (or 1 if the $c$-ghost is not taken into account). In the NS sector they correspond to the moduli $w$, $\bar{w}$, $a$ and $\chi$, and to the zero-momentum $A$ and $\phi$. We will focus on the four mixed directions, thus neglecting $\phi$ and $\chi$. We can join the remaining vertex operators into a matrix, taking into account all possible strings. This matrix has $(N+k)\times(N+k)$ entries, and is of the form (eventually rescaling the polarizations)
\begin{equation}\label{eq:vertex_psi}
V(z)=c(z) \begin{pmatrix}
V_A & V_w\\
V_{\bar{w}} & V_a
\end{pmatrix}(z)=\dfrac{g_{YM}}{\sqrt{\alpha'}} c(z)\begin{pmatrix}
A_\mu^{uv}\psi^\mu & w_{\dot{\alpha}}^{uj}\Delta S^{\dot{\alpha}}\\
\bar{w}_{\dot{\alpha}}^{iv}S^{\dot{\alpha}}\bar{\Delta} & a_\mu^{ij}\psi^\mu
\end{pmatrix}(z)e^{-\phi}(z)\,,
\end{equation}
where we have explicitly written the Chan-Paton indices. This vertex operator can be expressed in the canonical picture -1, as in \eqref{eq:vertex_psi}, or in picture 0. In order to change the picture of this vertex operator we notice that it has the form $V=c\mathbb{V}_{1/2}e^{-\phi}$, where $\mathbb{V}_{1/2}$ is a Grassmann-odd superconformal matter primary of weight $1/2$. Changing the picture on such a vertex operator is simple, and was discussed in \cite{Maccaferri:2018vwo}. Here we choose a slightly different notation: we call $X$ the picture changing operator, and we express it in terms of the BRST charge and the $\xi$ ghost as
\begin{equation}
X=\{Q,\xi\}\,.
\end{equation}
The BRST charge, in turn, is given explicitly by $Q=Q_0+Q_1+Q_2$ with (see \cite{Blumenhagen:2013fgp})
\begin{equation}\label{eq:Q}
\begin{split}
&Q_0=\oint \dfrac{dz}{2\pi i}\left(cT^{X,\psi,\beta,\gamma}+c(\partial c)b\right)(z)\,,\\
&Q_1=-\oint \dfrac{dz}{2\pi i}\,\gamma T_F(z)\,,\\
&Q_2=-\dfrac{1}{4}\oint \dfrac{dz}{2\pi i}\,b\gamma^2(z)\,,
\end{split}
\end{equation}
where $T_F$ is the matter supercurrent $T_F(z)=\frac{i}{\sqrt{2\alpha'}}\psi_\mu\partial X^\mu$. We then have 
\begin{equation}\label{XV}
    X\,V=-c\mathbb{V}_1+\frac{1}{4}\gamma\mathbb{V}_{1/2}\,,
\end{equation}
where $\mathbb{V}_1$ is a Grassmann-even superconformal primary of weight one defined by
\begin{equation}\label{eq:pic_changing}
T_F(z)\mathbb{V}_{1/2}(0)=\dfrac{\mathbb{V}_1(0)}{z}+\text{regular}\,.
\end{equation}
The explicit calculation for the vertex operator \eqref{eq:vertex_psi} gives
\begin{equation}\label{eq:vertex_psi_0}
\begin{split}
XV(z)= -\dfrac{g_{YM}}{\sqrt{\alpha'}}c\begin{pmatrix}
\dfrac{i}{\sqrt{2\alpha'}}A_\mu^{uv}\partial X^\mu & -\dfrac{1}{2\sqrt{2}}w_{\dot{\alpha}}^{uj}\tau_\mu(\bar{\sigma}^\mu)^{\dot{\alpha}}_{\;\;\beta}S^{\beta}\\
-\dfrac{1}{2\sqrt{2}}\bar{w}_{\dot{\alpha}}^{iv} (\bar{\sigma}^\mu)^{\dot{\alpha}}_{\;\;\beta}S^{\beta}\bar{\tau}_\mu & \dfrac{i}{\sqrt{2\alpha'}}a_\mu^{ij}\partial X^\mu
\end{pmatrix}(z)&+\\
+\dfrac{1}{4}\dfrac{g_{YM}}{\sqrt{\alpha'}}\gamma\begin{pmatrix}
A_\mu^{uv}\psi^\mu & w_{\dot{\alpha}}^{uj}\Delta S^{\dot{\alpha}}\\
\bar{w}_{\dot{\alpha}}^{iv}S^{\dot{\alpha}}\bar{\Delta} & a_\mu^{ij}\psi^\mu
\end{pmatrix}&(z)\,,
\end{split}
\end{equation}
where $\tau_\mu$ (and analogously $\bar{\tau}_\mu$) is a combination of an excited twist field along the $\mu$ direction and three normal twist fields along the other directions \cite{Mattiello:2018kue}.

\section{Second Order Deformation}\label{sec:SFT}
In this section we start analyzing exact marginality of the blow up of the size of a D(-1) brane in a D3 background. We will do this in the framework of super string field theory (SFT); this is necessary for two reasons: As we shall see shortly, we will encounter amplitudes with on-shell internal states as well as off-shell external states. The Yang-Mills action \eqref{eq:Effective4} arises, in the limit $\alpha'\to 0$, from open superstring field theory after integrating out massive and auxiliary fields \cite{Berkovits:2003ny,Maccaferri:2018vwo}. We will not need all of the technical material that goes into its construction. Let us instead begin by reviewing some details relevant for this paper.

\subsection{Open Superstring Field Theory} 
The NS sector of open superstring field theory (OSFT) is defined perturbatively on the space of states $\mathcal{H}$ of the worldsheet SCFT of $(-1)/(-1)$, $(-1)/3$, $3/(-1)$ and $3/3$ strings with a non-degenerate BPZ inner product 
\begin{equation}
\left(\Psi_1,\Psi_2\right)= \lim_{z\to 0}\text{Tr}\left\langle (I^\ast\mathcal{O}_{\Psi_1})(z)  \mathcal{O}_{\Psi_2}(z) \right\rangle_H\;\text{,}
\end{equation}
where the trace is over NN and DD boundary conditions, $\langle\dots\rangle_H$ is the correlator on the upper half plane and $I(z)=-1/z$ while $I^\ast\mathcal{O}$ denotes the conformal transformation of $\mathcal{O}$ with respect to $I$. The BPZ  inner product is graded symmetric due to the 3 ghost insertions originating from the $SL(2,\mathbb{R})$ isometry group of the disk. With this, the kinetic term is given by
\begin{equation}
\frac{1}{2}\left(\Psi,Q\Psi\right)\,,
\end{equation}
where $Q$ is the open string BRST charge of ghost number one and $\Psi$ is an arbitrary state in in the state space of the matter plus ghost SCFT. 

In addition to the quadratic term, OSFT has an infinite number of higher order interaction terms:
\begin{equation}\label{OSFT1}
S(\Psi)=\frac12\left(\Psi,Q\Psi\right)+\frac{1}{3}\left(\Psi,M_2(\Psi,\Psi)\right)+ \frac{1}{4}\left(\Psi,M_3(\Psi,\Psi,\Psi)\right)+ \cdots
\end{equation}
All of these vertices are contact terms, meaning that they do not involve integrals over even directions in moduli space. However, they do involve integrals over the odd directions, which are implemented by the insertion of a BPZ-even picture changing operator $X$ \cite{Erler:2013xta}. Let us focus on $M_2$ at present. Ignoring picture changing for the moment, $M_2$ reduces to an associative product
\begin{equation} 
m_2\;: \mathcal{H}\times \mathcal{H}\to \mathcal{H}\,,
\end{equation}
which can be described by the three-point correlator 
\begin{equation}\label{star}
\left(\Psi_1,m_2(\Psi_2,\Psi_3)\right):=\text{Tr}\left\langle (f_\infty^\ast\mathcal{O}_{\Psi_1})(0)(f_1^\ast\mathcal{O}_{\Psi_2})(0)(f_0^\ast\mathcal{O}_{\Psi_3})(0)\right\rangle\,,
\end{equation}
where $f_w(z)$ is a family of conformal maps from the half disk to the upper half plane such that $f_w(0)=w$. For now we will not need any further information on $f_w(z)$ since we will consider only conformal scalars fields, i.e. on-shell (except in section \ref{subs:profile}).  Equivalently, $m_2$ is defined in terms of the operator product expansion (OPE) of conformal fields. This will sometimes be more convenient for our use below. 

In order to implement the integration over the odd moduli we  define the picture changing operators \cite{Erler:2013xta}
\begin{equation}\label{defX}
X=\frac{1}{2\pi i}\oint\frac{dx}{z} X(z)\,\;,\quad \xi=\frac{1}{2\pi i}\oint\frac{dx}{z} \xi(z)
\end{equation}
around each puncture, with $X(z)=\{Q,\xi(z)\}$ where $\xi(z)$ is defined in \eqref{eq:superghosts}. The product $M_2$ can then be expressed as 
\begin{equation}\label{defM2}
M_2(A,B)=\frac{1}{3}\Big(Xm_2(A,B)+ m_2(XA,B)+m_2(A,XB)\Big)\,,
\end{equation}
where $Xm_2(A,B)$ can be evaluated with the help of the BPZ inner product, using 
\begin{equation}
(C,Xm_2(A,B))=(XC,m_2(A,B))\,.
\end{equation}
Note that $M_2$ so defined is associative only up to homotopy, that is a $Q$-exact term, due to the fact that $X$ does not commute with the $m_2$ operation\footnote{Note, that in contrast to \cite{Erler:2013xta}, here we do not work with the shifted (suspended) Hilbert space. Consequently some extra minus signs appear relative to \cite{Erler:2013xta} (see also appendix \ref{Pder}).}. Consequently the algebraic structure of OSFT is that of a homotopy associative (or $A_\infty$) algebra. This structure then uniquely determines the higher order products, up to field redefinitions. This is how the $A_\infty$-OSFT of \cite{Erler:2013xta} is constructed. However, we will not need any details of this construction other than the fact that $M_3$ cancels the non-associativity of $M_2$ and that $M_2$ itself is exact in the large Hilbert space, that is, 
\begin{equation}\label{mu}
    M_2=\{Q,\mu_2\}\,,\qquad\mu_2(A,B)=\frac{1}{3}\left(\xi m_2(A,B)+ m_2(\xi A,B)+(-1)^{|A|}m_2(A,\xi B)\right)\,.
\end{equation}

\subsection{Second Order Deformation}\label{sub:exact_marginality}
A marginal deformation in the worldsheet CFT is exactly marginal if the corresponding solution of the linearized equation of motion of OSFT can be integrated to a solution to the nonlinear equation of motion. So let us  start by writing down the equation of motion following from \eqref{OSFT1}
\begin{equation}\label{eomS}
Q\Psi+M_2(\Psi,\Psi)+M_3(\Psi,\Psi,\Psi)+\cdots =0\,.
\end{equation}
We then expand the field in a perturbation series with parameter $\frac{\rho}{\sqrt{\alpha'}}$:
\begin{equation}
\Psi=\dfrac{\rho}{\sqrt{\alpha'}}\Psi_0+\left(\dfrac{\rho}{\sqrt{\alpha'}}\right)^2\Psi_1+\left(\dfrac{\rho}{\sqrt{\alpha'}}\right)^3\Psi_2+\dots\,,
\end{equation}
where
\begin{equation}\label{psi-}
\dfrac{\rho}{\sqrt{\alpha'}}\Psi_0=V\,,
\end{equation}
with $V$ as in \eqref{eq:vertex_psi}, with moduli $w$, $\bar{w}$ and $a$ in $\psi_0$ satisfying the ADHM constraints. This is  a solution of the linearized equation of motion $QV=0$ that describes an infinitesimal blow-up of the D(-1) brane, and $(\frac{\rho}{\sqrt{\alpha'}})^2\Psi_1 +\dots$ are the higher order corrections to $(\frac{\rho}{\sqrt{\alpha'}})\Psi_0$.  To first order in the non-linearity (second order in $\rho/\sqrt{\alpha'}$) we then have 
\begin{equation}
Q\Psi_1+M_2(\Psi_0,\Psi_0)=0\,,
\end{equation}
which is solved by 
\begin{equation}\label{sol1}
\Psi_1=-Q^{-1}M_2(\Psi_0,\Psi_0)+\psi_1\,,
\end{equation}
with $\psi_1$ a solution to the homogeneous equation, $Q\psi_1=0$. Equation \eqref{sol1} is well defined provided that $Q$ has an inverse. For this we need to choose a gauge fixing. Here we will work in Siegel gauge, $b_0\Psi=0$, with $Q^{-1}=\frac{b_0}{L_0}$. Then
\begin{equation}\label{qqm}
Q\,Q^{-1}+Q^{-1}\,Q=1-P_0\,,
\end{equation}
where $P_0$ is the projector on the cohomology $H(Q)\subset \text{ker}(L_0)$, satisfying
\begin{equation}
    QP_0=P_0Q=0\,.
\end{equation}
To see if $\Psi_1$ in \eqref{sol1} is well defined we then compute
\begin{equation}
Q\Psi_1=-Q\,Q^{-1}M_2(\Psi_0,\Psi_0)=(Q^{-1}Q+P_0-1) M_2(\Psi_0,\Psi_0)\,.
\end{equation}
The first term on the r.h.s vanishes using $[Q,M_2]=0$ (which, in turn, follows from the fact that $[Q,X]=0$) and that $Q$ is a derivation of the star product $m_2$ defined through \eqref{star}. Thus \eqref{sol1} is meaningful provided that
\begin{equation}\label{PM2}
P_0 M_2(\Psi_0,\Psi_0)=0.
\end{equation}
To prove that \eqref{PM2} holds we show that $P_0M_2(V,V)=0$ for a vertex operator of the form $V(z)=c\mathbb{V}_{1/2}\,e^{-\phi}(z)$, where $\mathbb{V}_{1/2}$ is a matter vertex operator of conformal dimension $1/2$; this is the prototype of vertex operator we are interested in (cfr. \eqref{eq:vertex_psi}). Let us first consider the contribution  $P_0[m_2(V,XV)+m_2(XV,V)]$ in \eqref{defM2}. Since all operators involved have total conformal weight zero we can evaluate this expression using the OPE. That is, using \eqref{XV},
\begin{equation}
\begin{split}
&P_0[m_2(V,XV)+m_2(XV,V)]=\lim_{z\rightarrow 0}P_0\left[V(z)XV(0)+XV(0)V(z)\right]=\\
&=\lim_{z\rightarrow 0}P_0\left[c \mathbb{V}_{1/2}e^{-\phi}(z)\left(-c \mathbb{V}_1(0)+\dfrac{1}{4}\gamma \mathbb{V}_{1/2}(0)\right)+\left(-c \mathbb{V}_1(z)+\dfrac{1}{4}\gamma \mathbb{V}_{1/2}(z)\right)c \mathbb{V}_{1/2}e^{-\phi}(0)\right]\,,
\end{split}
\end{equation}
where $\mathbb{V}_1$ is the matter operator of conformal dimension $1$ defined in \eqref{eq:pic_changing}. Since the OPE $\mathbb{V}_{1/2}(z)\mathbb{V}_{1/2}(0)$ contains a single pole, while the OPE $\mathbb{V}_{1/2}(z)\mathbb{V}_{1}(0)$ does not, we conclude that\footnote{We would like to thank C. Maccaferri for helpful comments on this point.}
\begin{equation}
P_0[m_2(V,XV)+m_2(XV,V)]=\lim_{z\rightarrow 0}P_0\left[\dfrac{1}{4} z\,c\eta(0)\left(\mathbb{V}_{1/2}(z)\mathbb{V}_{1/2}(0)-\mathbb{V}_{1/2}(z)\mathbb{V}_{1/2}(0)\right)\right]=0\,,
\end{equation}
where we used the fact that $\mathbb{V}_{1/2}$ and $c$ are fermionic operators. Let us then consider the remaining term $P_0[Xm_2(V,V)]=XP_0[m_2(V,V)]$ in \eqref{defM2}. Recalling \eqref{qqm} we restrict the OPE to the kernel of $L_0$, 
\begin{equation}\label{eq:m2VV}
m_2(V,V)\vert_{\text{ker}(L_0)}=\lim_{z\rightarrow 0}(c\mathbb{V}_{1/2}\,e^{-\phi})(z)(c\mathbb{V}_{1/2}\,e^{-\phi})(0)=\partial(c\partial c e^{-2\phi}\mathbb{V}'_0)+ c\partial c\mathbb{V}'_1 e^{-2\phi}\,,
\end{equation}
where $\mathbb{V}'_0$ and $\mathbb{V}'_1$ are matter vertex operators of conformal weight 0 (thus proportional to the identity) and 1 respectively. It is not hard to see that the first term in \eqref{eq:m2VV} is $Q$-exact, i.e. 
\begin{equation}
    \partial(c\partial c e^{-2\phi}\mathbb{V}'_0)=Q\left(\partial c e^{-2\phi}\mathbb{V}'_0\right)\,,
\end{equation}
and therefore it is annihilated by $P_0$, since $P_0Q=0$, leaving
\begin{equation}
    P_0m_2(V,V)= c\partial c\mathbb{V}'_1 e^{-2\phi}\,.
\end{equation}
For the second term in \eqref{eq:m2VV} we proceed using the identity  $X=\{Q,\xi\}$ in the large Hilbert space. Since $V$ is on-shell, 
\begin{equation}\label{eq:Qfinal}
P_0 M_2(V,V)=Q\left(\xi c\partial c\mathbb{V}'_1 e^{-2\phi}\right)=(Q_0+Q_1+Q_2)
\left(\xi c\partial c\mathbb{V}'_1 e^{-2\phi}\right)\,,
\end{equation}
where we used the explicit definition \eqref{eq:Q}. Bosonizing the ghosts as in appendix \ref{app:OPE} we can compute each term in \eqref{eq:Qfinal}. The first and last term clearly  vanish, while the second term extracts the double pole of the OPE
\begin{equation}
T_F(z)\mathbb{V}'_1(0)\,.
\end{equation}
In our case, however, the operator $\mathbb{V}'_1$ is proportional to $\psi^{\mu\nu}$, as we will see in section \ref{sec:third_order}. The OPE with the supercurrent is then given by
\begin{equation}
T_F(z)\psi^{\mu\nu}(0)\sim \dfrac{1}{z}\left(\partial X^\mu\psi^\nu-\partial X^\nu\psi^\mu \right)(0)+\dots
\end{equation}
Therefore we conclude that \eqref{eq:Qfinal} vanishes, thus establishing that  $P_0M_2(V,V)=0$. Hence, the first order correction $\Psi_1$ in \eqref{sol1} is well defined, even without specifying the ADHM constraints.

\section{Instanton Profile}\label{subs:profile}
As an application we compute the first order contribution to the instanton profile, that is the projection of $\Psi_1$ to a gluon state. This should correspond to computing the instanton profile at second order in $\rho/\sqrt{\alpha'}$, valid for  $\rho^2\ll\alpha'$. For simplicity we set $a_\mu=0$; a different value for $a_\mu$ would correspond to moving the position of the instanton. 
Concretely we consider
\begin{equation}\label{eq:profile}
A_\mu^{c(1)}=\left(\dfrac{\rho}{\sqrt{\alpha'}}\right)^2(\mathcal{V}_{A_\mu^c},\Psi_1)=-\left(\dfrac{\rho}{\sqrt{\alpha'}}\right)^2(\mathcal{V}_{A_\mu^c},Q^{-1}M_2(\Psi_0,\Psi_0))\,.
\end{equation}
Since $\mathcal{V}_{A_\mu^c}$ is a vertex operator in the 3/3 sector, this matrix element projects the 3/3 component of $M_2(\Psi_0,\Psi_0)$. Thus 
\begin{equation}\label{eq:profilep}
A_\mu^{c(1)}=-(\mathcal{V}_{A_\mu^c},Q^{-1}M_2(V_w,V_{\bar{w}}))\,,
\end{equation}
where we used the same symbol $M_2$ for the matrix 
components of $M_2$. The latter involves picture changing operators on the inputs as well as on the output of the product. However, since $X$ is a conformal scalar we can pull it through the propagator $Q^{-1}$ onto $\mathcal{V}_{A_\mu^c}$. Furthermore since none of the the vertex operators involve the $\eta$ ghost, we can move $X$ from either input to $\mathcal{V}_{A_\mu^c}$ in spite of $\mathcal{V}_{A_\mu^c}$ being off-shell. Consequently we can take $\mathcal{V}_{A_\mu^c}$ in picture zero while the boundary changing vertex operators are in picture -1.

\begin{figure}[ht]
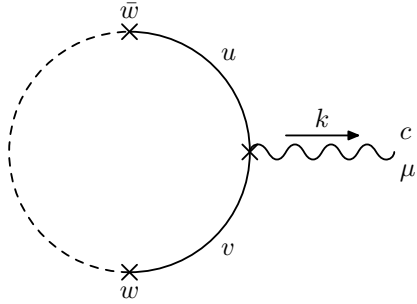

\centering
$\hbox{ \convertMPtoPDF{profile.1}{1}{1} }$
\caption{First order contribution to the instanton profile. The solid line represents the D3 branes, while the dashed one represents the D(-1) brane; the indices $u,v=1,2$ label the particular D3 branes. The vector $A_\mu^c$ (with outgoing momentum $k^\nu$) comes from a 3/3 string, and the corresponding vertex operator has to be inserted in the middle of the solid line. The curly line represents the presence of a gluon propagator.}
\label{fig:First_Order}
\end{figure}

The calculation of this quantity can be done in two steps. Using the definition \eqref{star}, we first need to compute the correlator
\begin{equation}\label{eq:profile_1}
\text{Tr}\langle\langle (f^*_\infty V_{\bar{w}}^{(-1)})(0)(f^*_1\mathcal{V}_{A_\mu^c}^{(0)})(0;-k)(f^*_0 V_{w}^{(-1)})(0)\rangle\rangle_{D(-1)}\,,
\end{equation}
where $V_{\bar{w}}$ and $V_w$ are boundary changing operators, and $\mathcal{V}_{A_\mu^c}$ is the gluon vertex operator with outgoing momentum, with a free Lorentz  and color index. Notice that the topological normalization is the one of the lowest brane \cite{Billo:2002hm}. Then we act with $Q^{-1}$, which, in Siegel gauge, results in multiplication by $1/k^2$. The calculation of \eqref{eq:profile} is sketched in figure \ref{fig:First_Order}. 
\\
Explicitly, the boundary changing operators (in picture -1) in \eqref{eq:profile_1} are the ones given in \eqref{eq:vertex_3-1_NS} with the rescaling \eqref{eq:rescaling}, while $\mathcal{V}_{A_\mu}$ is given (in picture -1) by
\begin{equation}\label{eq:gluon_v-1}
\mathcal{V}_{A_\mu^c}^{(-1)uv}(z;-k)=\sqrt{\dfrac{\alpha'}{2}}\dfrac{(\tau^c)^{uv}}{2}c(z)\psi_\mu(z)e^{-\phi(z)}e^{-ik\cdot X(z)}\,,
\end{equation}
where we have used the Chan-Paton factor $(T^c)^{uv}=(\tau^c)^{uv}/2i$. Applying the picture changing operator to \eqref{eq:gluon_v-1} we get
\begin{equation}\label{eq:gluon_vo}
\mathcal{V}^{(0)uv}_{A^c_{\mu}}(z;-k)=\dfrac{(\tau^c)^{vu}}{2}(i\partial X_\mu-\dfrac{\alpha'}{2} k\cdot \psi \psi_\mu)e^{-ik\cdot X(z)}\,.
\end{equation}
In contrast to \eqref{eq:vertex_psi_0}, there is an extra contribution due to the non-vanishing momentum $k^\mu$. Furthermore, only the term with a $c$-ghost (and not the one with a $\gamma$-ghost) can contribute to the correlation function \eqref{eq:profile_1}.
We note here that the action of the maps $f_z$ reduces for primary operator $A(w)$ to
\begin{equation}
 f_z^* A(w)= f_z'(w)^{h}A(f_z(w))\,,
\end{equation} 
where $h$ is the conformal dimension.
In our case, since two operators are on-shell, we need only one map $f_1$, for the gauge vector. Therefore the correlation function reduces to
\begin{equation}
A_{\mu}^{c(1)}(k)=C_0 f_1'(0)^{\alpha'k^2/2}\langle V_{\bar{w}}^{(-1)u}(\infty)\mathcal{V}^{(0)uv}_{A_{\mu}}(1;k)V_{w}^{(-1)v}(0) \rangle\,,
\end{equation}
where we made the $SU(2)$ indices explicit. The detailed calculation is done in appendix \ref{app:profile}. Here we state the final result in momentum space, assuming the ADHM constraints, that is
\begin{equation}\label{eq:instanton_1_momentum_main}
A_{\mu}^{c(1)}(k)=\left(\dfrac{f_1'(0)}{4}\right)^{\alpha'k^2/2}i\rho^2 k^\nu\bar{\eta}^c_{\nu\mu}e^{-ik\cdot x_0}\,,
\end{equation}
where the factor $(1/4)^{\alpha'k^2/2}$ takes into account the proper normal ordering on a twisted background \cite{Mukhopadhyay:2001ey,Pesando:2011yd}\footnote{We would like to thank Igor Pesando for clarifying comments on this point.}. Notice that this result depends on $\alpha'$ and on the choice of the map $f_1$ (different maps correspond to different field redefinitions); \eqref{eq:instanton_1_momentum_main} differs from the result in \cite{Billo:2002hm}, where the off-shell amplitude \eqref{eq:profile_1} was computed within the on-shell formalism. Let us now perform a Fourier transform, in order to have a result in configuration space; as explained above, the propagator in Siegel gauge has to be added. The result is (see appendix \ref{app:profile} for the detailed calculation)
\begin{equation}\label{profalpha}
A_{\mu}^{c\,(1)}(x)=2\rho^2\bar{\eta}^c_{\mu\nu}\dfrac{(x-x_0)^\nu}{(x-x_0)^4}\left[1+e^{(x-x_0)^2/(2\alpha' L_1)}\left(1-\dfrac{(x-x_0)^2}{2\alpha'L_1}\right)\right]\,,
\end{equation}
where $L_1=\log (f'_1(0)/4)<0$ since $f'_1(0)<1$. In the field theory limit $\alpha' k^2\ll 1$ the dependence on $f'_1(0)$ drops out. Since we also assumed from the beginning that $\rho^2\ll\alpha'$, the field theory limit will also correspond to a large distance (compared to the size $\rho$) limit. In this limit the profile in position space is (see appendix \ref{app:profile})
\begin{equation}
A_{\mu}^{c\,(1)}(x)=2\rho^2\bar{\eta}^c_{\mu\nu}\dfrac{(x-x_0)^\nu}{(x-x_0)^4}\,,
\end{equation}
which is exactly the leading term  in a large distance expansion ($\rho^2\ll(x-x_0)^2$) of the full $SU(2)$ instanton solution \eqref{eq:singular}, as previously found in \cite{Billo:2002hm}.

In closing this section we note that a zero momentum gluon, appearing in vertex operator in \eqref{eq:vertex_psi}, does not source a non-linear correction to (\ref{profalpha}). This is because the correction would be proportional to the three point function
\begin{equation}
    \langle V_{A}^{(-1)}(\infty;0)\mathcal{V}^{(0)}_{A}(1;k)V_{A}^{(-1)}(0;0) \rangle\,,
\end{equation}
which vanishes, since two of the vertex operators have vanishing momentum. Thus,  the complete profile up to order $\rho^2$ is given by
\begin{equation}\label{eq:full_profile}
    A_\mu^c(x)= A_\mu^{c\,(0)}+A_{\mu}^{c\,(1)}(x)\,,
\end{equation}
where $A_\mu^{c\,(0)}$ is constant in position space and $A_{\mu}^{c\,(1)}(x)$ given by (\ref{profalpha}). For the same reason this zero momentum gluon does not source a deformation in the  3/(-1), (-1)/3 or (-1)/(-1) sectors.

\section{Third Order Deformation}\label{sec:third_order}
At second order in the deformation (third order in $\rho/\sqrt{\alpha'}$) the equation of motion \eqref{eomS} reads
\begin{equation}\label{eq:psi2}
Q\Psi_2-M_2(Q^{-1}M_2(\Psi_0,\Psi_0)-\psi_1,\Psi_0)-M_2(\Psi_0,Q^{-1}M_2(\Psi_0,\Psi_0)-\psi_1)+M_3(\Psi_0,\Psi_0,\Psi_0) =0\,,
\end{equation}
where we used the solution for $\Psi_1$ given in \eqref{sol1}. The obstruction to inverting $Q$ is given by 
\begin{equation}
\begin{split}
(Q^{-1}Q+P_0)\Big[M_2(Q^{-1}M_2(\Psi_0,&\Psi_0)-\psi_1,\Psi_0)+\\
&+M_2(\Psi_0,Q^{-1}M_2(\Psi_0,\Psi_0)-\psi_1)-M_3(\Psi_0,\Psi_0,\Psi_0)\Big]\,.
\end{split}
\end{equation}
Let us first consider the terms involving $Q^{-1}Q$. They add up to (using also $Q\psi_1=0$)
\begin{equation}\label{o2nd1}
    Q^{-1}\left[M_2(M_2(\Psi_0,\Psi_0),\Psi_0)-M_2(\Psi_0,M_2(\Psi_0,\Psi_0))-QM_3(\Psi_0,\Psi_0,\Psi_0)\right]\,,
\end{equation}
which vanishes by the $A_\infty$ relations (see e.g \cite{Erler:2013xta,Moeller:2010mh}). The remaining obstruction is then 
\begin{equation}\label{p02}
P_0\left[ M_2(Q^{-1}M_2(\Psi_0,\Psi_0)-\psi_1,\Psi_0)+M_2(\Psi_0,Q^{-1}M_2(\Psi_0,\Psi_0)-\psi_1)-M_3(\Psi_0,\Psi_0,\Psi_0)\right]\,.
\end{equation}
We note, in passing, that \eqref{p02} is just the minimal model map to fourth order of the underlying $A_\infty$ algebra which extracts the $S$-matrix elements of string field theory. This does not come as a surprise, since $S$-matrix elements are known to be given by the obstructions of a linearized solution to give rise to a non-linear solution (e.g. \cite{Konopka:2015tta}). 

In order to analyze this obstruction we first note that $P_0 M_2(\psi_1,\Psi_0)$ and $P_0 M_2(\Psi_0,\psi_1)$ vanish. The proof of this is completely analogous to that given above for  $P_0M_2(\Psi_0,\Psi_0)$. Next, we consider $P_0 M_2(Q^{-1}M_2(\Psi_0,\Psi_0),\Psi_0)$, which we write as 
\begin{equation}
\sum\limits_ie^i\langle e_i,M_2(Q^{-1}M_2(\Psi_0,\Psi_0),\Psi_0)\rangle\,,
\end{equation}
where $e_i$ ($e^i$) is a basis (and its dual) of the image of $P_0$ with $\langle e^i,e_j\rangle=\delta^i_j$. To continue we use \eqref{mu} to write
\begin{equation}\label{red1}
\begin{split}
\langle e_i,M_2(Q^{-1}M_2(\Psi_0,\Psi_0),\Psi_0)\rangle&= -\frac12\langle e_i,\xi\;M_2(Q^{-1}\{Q,\mu_2\}(\Psi_0,\Psi_0),\Psi_0)\rangle_L+\\
&- \frac12\langle e_i,\xi\;\{Q, \mu_2\}(Q^{-1}M_2(\Psi_0,\Psi_0),\Psi_0)\rangle_L\,,
\end{split}
\end{equation}
where, since \eqref{mu} holds only in the large Hilbert space $H_L$, we now use the BPZ inner product in $H_L$  with an extra insertion of $\xi$ to saturate the extra zero mode in $H_L$. The second term in \eqref{p02} is treated analogously. Next we commute $Q^{-1}$ across $Q$ and use the fact that $Q$ commutes with $M_2$ and annihilates $\Psi_0$.
In doing so we pick up the contributions 
\begin{equation}\label{red2}
\begin{split}
& \frac12\langle e_i,\xi\;M_2(P_0\mu_2(\Psi_0,\Psi_0),\Psi_0)\rangle_L+ \frac12\langle e_i,X\,M_2(Q^{-1}\mu_2(\Psi_0,\Psi_0),\Psi_0)\rangle_L+\\
&- \frac12\langle e_i,X\, \mu_2(Q^{-1}M_2(\Psi_0,\Psi_0),\Psi_0)\rangle_L-\frac12\langle e_i,\xi\, \mu_2(P_0M_2(\Psi_0,\Psi_0),\Psi_0)\rangle_L \,.
\end{split}
\end{equation}
The last term above vanishes for the same reason as in subsection \ref{sub:exact_marginality}. 
In the two terms above involving $X$, the $\xi$ zero-mode must be provided by $\mu_2$, so that, for instance, 
\begin{equation}\label{midr}
-\frac12\langle e_i,X\, \mu_2(Q^{-1}M_2(\Psi_0,\Psi_0),\Psi_0)\rangle_L= -\frac12\langle e_i,X\xi\, m_2(Q^{-1}M_2(\Psi_0,\Psi_0),\Psi_0)\rangle_L
\end{equation}
and similarly for the second term in (\ref{red2}). In what follows, we will neglect the terms that originate from the identity in \eqref{qqm} since, as shown in \cite{Erler:2013xta}, these cancel against $M_3$ in \eqref{p02}. Applying \eqref{mu} to \eqref{midr} we get
\begin{equation}\label{red3}
-\frac12\langle e_i,X\, \mu_2(Q^{-1}M_2(\Psi_0,\Psi_0),\Psi_0)\rangle_L= -\frac12\langle e_i,X\xi\, m_2(Q^{-1}\{Q,\mu_2\}(\Psi_0,\Psi_0),\Psi_0)\rangle_L\,.
\end{equation}
We then commute $Q^{-1}$ across $Q$ and use again that $Q$ annihilates $\Psi_0$. Therefore \eqref{red3} gives a contribution 
\begin{equation}\label{red4}
\frac12\langle e_i,X\xi\, m_2(P_0\,\mu_2(\Psi_0,\Psi_0),\Psi_0)\rangle_L-\frac12\langle e_i,X^2\, m_2(Q^{-1}\,\mu_2(\Psi_0,\Psi_0),\Psi_0)\rangle_L\,.
\end{equation}
The objective of deriving the expressions \eqref{red2} and \eqref{red4} was to isolate the contact terms that originate in the integration over odd moduli in the super moduli space (encoded in the super string products $M_2$ and $M_3$). This procedure can be applied in complete analogy to the remaining terms in \eqref{p02}. More details on this derivation can be found in appendix \ref{Pder}. 
The result is a sum of two contributions, one involving the projector $P_0$ and the other involving the propagator $Q^{-1}$. The first contribution reads\footnote{We would like to thank Jakub Vo\v{s}mera four pointing out a sign mistake in this equation.}
\begin{align}\label{P02b}
   &A= -\frac{1}{3}\langle P_0\mu_2(\Psi_0,\Psi_0), 4m_2(\Psi_0,\xi Xe_i)-4m_2(\xi Xe_i,\Psi_0) +  m_2(X\Psi_0,\xi e_i)-m_2(\xi e_i,X\Psi_0) \rangle_L\nonumber\\
   &-\frac{1}{3}\langle P_0m_2(\Psi_0,\Psi_0),\xi m_2(\Psi_0, \xi Xe_i)-\xi m_2(\xi X e_i,\Psi_0)+m_2(\xi \Psi_0, \xi Xe_i)- m_2(\xi X e_i, \xi \Psi_0)\rangle_L\nonumber\\
    &-\frac{1}{3}\langle XP_0\mu_2(\Psi_0,\Psi_0), m_2(\Psi_0,\xi e_i) - m_2(\xi e_i,\Psi_0)\rangle_L\,,
\end{align}
where we used the cyclic properties of the string products $m_2$ and $\mu_2$ (see Appendix \ref{Pder}) as well as $X\xi \Psi_0=\xi X\Psi_0$ (and similarly for $e_i$). The second term, involving the propagator is given by 
\begin{equation}\label{redwsp}
B=-2\langle X\circ X\,e_i, m_2( Q^{-1}m_2(\Psi_0,\Psi_0),\Psi_0) + m_2(\Psi_0,Q^{-1} m_2(\Psi_0, \Psi_0)\rangle\,.
\end{equation}
In the next two subsections we will evaluate these two terms separately.  

\subsection{Evaluation of A}
To continue we evaluate the terms appearing in \eqref{P02b}. In principle there are anomalous contributions due to the fact that $\xi X e_i$ contains the operator $:\xi\eta:$ which is not primary; we are going to discuss this problem in appendix \ref{app:anomaly}, where we show that all anomalous contributions cancel. For the moment we proceed as we would do if all the vertex operators were primaries; let us start with $P_0 m_2(\Psi_0,\Psi_0)$.  Using the OPE relations given in appendix \ref{app:OPE} with $\frac{\rho}{\sqrt{\alpha'}}\Psi_0=V$, we find (for a single D($-$1) brane, i.e. $k=1$ and assuming $a_\mu=0$ for simplicity)
\begin{equation}\label{eq:p0m2_a}
\left(\dfrac{\rho}{\sqrt{\alpha'}}\right)^2P_0m_2(\Psi_0,\Psi_0)=P_0m_2(V,V)=\dfrac{1}{2}\dfrac{g^2_{YM}}{\alpha'} c\partial c e^{-2\phi}\psi^{\mu\nu}M_{\mu\nu}\,,
\end{equation}
with
\begin{equation}\label{eq:Mmunu}
M_{\mu\nu}=\begin{pmatrix}
[A_\mu,A_\nu]+\frac{1}{2}w_{\dot{\alpha}}(\bar{\sigma}_{\mu\nu})^{\dot{\alpha}\dot{\beta}}\bar{w}_{\dot{\beta}} & 0\\
0 & \frac{1}{2}\bar{w}_{\dot{\alpha}}(\bar{\sigma}_{\mu\nu})^{\dot{\alpha}\dot{\beta}}w_{\dot{\beta}}
\end{pmatrix}\,,
\end{equation}
where we have projected out the $Q$-exact piece, in analogy to the one in \eqref{eq:m2VV}. On the other hand,   $P_0\mu_2(V,V)$ is given by
\begin{equation}\label{eq:p0mu2}
\left(\dfrac{\rho}{\sqrt{\alpha'}}\right)^2P_0\mu_2(\Psi_0,\Psi_0)=P_0\mu_2(V,V)=\dfrac{1}{2}\dfrac{g^2_{YM}}{\alpha'} \xi c\partial c e^{-2\phi}\psi^{\mu\nu}M_{\mu\nu}
+ 
\dfrac{g^2_{YM}}{3\alpha'} \partial \xi c\partial c e^{-2\phi}U\,,
\end{equation}
where the last term is proportional to the identity in the matter sector and
\begin{equation}
    U=\begin{pmatrix}
A_\mu A_\nu\delta^{\mu\nu}+w_{\dot{\alpha}}\epsilon^{\dot{\alpha}\dot{\beta}}\bar{w}_{\dot{\beta}} & 0\\
0 &  \bar{w}_{\dot{\alpha}}\epsilon^{\dot{\alpha}\dot{\beta}}w_{\dot{\beta}}
\end{pmatrix}\,.
\end{equation}
To continue we note that, without restricting the generality, we may parametrize a generic zero-momentum Siegel gauge state $e_i$ in physical subspace $H_{phys}$  by 
\begin{equation}\label{eq:e_i}
e_i(z)= \dfrac{g_{YM}}{\sqrt{\alpha'}}c(z)\begin{pmatrix}
B_\mu\psi^\mu & v_{\dot{\alpha}}\Delta S^{\dot{\alpha}}\\
\bar{v}_{\dot{\alpha}}S^{\dot{\alpha}}\bar{\Delta} & b_\mu\psi^\mu
\end{pmatrix}(z)e^{-\phi}(z)\,,
\end{equation}
which is basically the same as \eqref{eq:vertex_psi}, but with a different generic polarizations $B_\mu$, $v_{\dot{\alpha}}$, $\bar{v}_{\dot{\alpha}}$ and $b_\mu$. In order to evaluate $A$ We need the explicit expressions of $\xi e_i$ and $\xi X e_i$, given by
\begin{equation}
\xi e_i= \dfrac{g_{YM}}{\sqrt{\alpha'}}\xi c\begin{pmatrix}
B_\mu\psi^\mu & v_{\dot{\alpha}}\Delta S^{\dot{\alpha}}\\
\bar{v}_{\dot{\alpha}}S^{\dot{\alpha}}\bar{\Delta} & b_\mu\psi^\mu
\end{pmatrix}e^{-\phi}\,,\qquad \xi X e_i= \dfrac{g_{YM}}{4\sqrt{\alpha'}}:\xi\eta:e^{\phi}\begin{pmatrix}
B_\mu\psi^\mu & v_{\dot{\alpha}}\Delta S^{\dot{\alpha}}\\
\bar{v}_{\dot{\alpha}}S^{\dot{\alpha}}\bar{\Delta} & b_\mu\psi^\mu
\end{pmatrix}\,,
\end{equation}
where we used \eqref{defX} and  for $Xe_i$ we kept only the term with the $\gamma$-ghost in \eqref{XV}. This is because all the terms involving $\xi X e_i$ in \eqref{P02b} already have three $c$-ghost insertions, therefore only the term with a $\gamma$ ghost in $\xi X e_i$ can contribute to the correlation functions. Using the OPE's in appendix \ref{app:OPE} we can check that 
\begin{equation}\label{eq:VxiXei}
P_0[m_2(V,\xi X e_i)-m_2(\xi X e_i,V)]=P_0[V,\xi X e_i]_{m_2}=\dfrac{g^2_{YM}}{4\alpha'}c:\xi\eta:W \,,
\end{equation}
with
\begin{equation}
    W=\begin{pmatrix}
A_\mu B_\nu\delta^{\mu\nu}+w_{\dot{\alpha}}\epsilon^{\dot{\alpha}\dot{\beta}}\bar{v}_{\dot{\beta}} & 0\\
0 &  \bar{w}_{\dot{\alpha}}\epsilon^{\dot{\alpha}\dot{\beta}}v_{\dot{\beta}}\,,
\end{pmatrix}-\begin{pmatrix}
B_\mu A_\nu\delta^{\mu\nu}+v_{\dot{\alpha}}\epsilon^{\dot{\alpha}\dot{\beta}}\bar{w}_{\dot{\beta}} & 0\\
0 &  \bar{v}_{\dot{\alpha}}\epsilon^{\dot{\alpha}\dot{\beta}}w_{\dot{\beta}}\,,
\end{pmatrix}\equiv W'-W''
\end{equation}
while
\begin{equation}\label{pxm}
\qquad P_0\xi m_2(V,\xi X e_i)=P_0\xi m_2(\xi X e_i,V)=0\,.
\end{equation}
The terms in \eqref{eq:VxiXei}, when coupled to \eqref{eq:p0m2_a}, cannot contribute to \eqref{P02b}, since they produce the one-point function $\langle\psi^{\mu\nu}\rangle$ in the matter sector, which vanishes. However, we get a non-vanishing contribution from the remaining terms in \eqref{P02b}. In particular, we can compute
\begin{equation}\label{op11}
\begin{split}
&
P_0[XV,\xi e_i]_{m_2}=\dfrac{g^2_{YM}}{4\alpha'} c\psi^{\rho\sigma}N_{\rho\sigma}-\dfrac{g^2_{YM}}{4\alpha'}c:\xi\eta:W+\dfrac{g^2_{YM}}{4\alpha'}c\partial\phi W+\dfrac{g^2_{YM}}{4\alpha'}\partial c W''\,,\\
&
P_0[\xi V,\xi X e_i]_{m_2}=-\dfrac{g^2_{YM}}{4\alpha'}\xi_0 c \psi^{\rho\sigma}N_{\rho\sigma}+\dfrac{g^2_{YM}}{4\alpha'}\xi_0 c\partial\phi W-\dfrac{g^2_{YM}}{4\alpha'}\xi_0\partial cW'-\dfrac{g^2_{YM}}{4\alpha'}\partial\xi cW''\,,\\
&
P_0[V,\xi e_i]_{m_2}=-\dfrac{g^2_{YM}}{\alpha'}\xi_0 c\partial c e^{-2\phi}\psi^{\rho\sigma}N_{\rho\sigma}-\dfrac{g^2_{YM}}{\alpha'}\partial\xi  c\partial c e^{-2\phi}W''\,,
\end{split}
\end{equation}
where $[\,\cdot,\cdot\,]_{m_2}$ denotes the graded commutator with respect to Witten's star product $m_2$, 
\begin{equation}\label{eq:Nrhosigma}
N_{\rho\sigma}=\begin{pmatrix}
[A_\rho,B_\sigma]+\frac{1}{4}\left(w_{\dot{\gamma}}(\bar{\sigma}_{\rho\sigma})^{\dot{\gamma}\dot{\delta}}\bar{v}_{\dot{\delta}}+ v_{\dot{\gamma}}(\bar{\sigma}_{\rho\sigma})^{\dot{\gamma}\dot{\delta}}\bar{w}_{\dot{\delta}} \right) & 0\\
0 & \frac{1}{4}\left(\bar{v}_{\dot{\gamma}}(\bar{\sigma}_{\rho\sigma})^{\dot{\gamma}\dot{\delta}}w_{\dot{\delta}}+v_{\dot{\gamma}}(\bar{\sigma}_{\rho\sigma})^{\dot{\gamma}\dot{\delta}}\bar{w}_{\dot{\delta}}   \right)
\end{pmatrix}\,,
\end{equation}
and $W$, $W'$ and $W''$ were defined above. Here $\xi_0$ is the zero mode of $\xi$. In the last line of \eqref{op11} we furthermore used that the combination
\begin{equation}
    \dfrac{g^2_{YM}}{\alpha'}\xi c e^{-2\phi}\left(\partial c\partial\phi -\frac{1}{2}\partial^2 c\right)(W'+W'')
\end{equation}
contributing to $[V,\xi e_i]_{m_2}$ is $Q$-exact and thus annihilated by the projector $P_0$. Indeed, $Q(\partial c e^{-2\phi})=c\partial^2ce^{-2\phi}-2\partial\phi c\partial c e^{-2\phi}$. Again, only the term with a $\gamma$ ghost in $XV$ can contribute to \eqref{P02b}, because it is inserted inside correlation functions with already three $c$-ghost insertions. Let us now contract the terms in \eqref{op11} with \eqref{eq:p0m2_a} and \eqref{eq:p0mu2} respectively. Focussing first on the terms containing the matter operator $\psi_{\mu\nu}$, and using the known correlation functions
\begin{equation}
\begin{gathered}
\langle \xi c\partial c e^{-2\phi}(z)c(w)\rangle_L=-(z-w)^2\,,\\
\langle \psi^{\mu\nu}(z)\psi^{\rho\sigma}(w)\rangle_L=\dfrac{-\delta^{\mu\rho}\delta^{\nu\sigma}+\delta^{\mu\sigma}\delta^{\nu\rho}}{(z-w)^2}\,,
\end{gathered}
\end{equation} 
we conclude that the first two lines of \eqref{P02b} exactly cancel for the state $e_i$. First, there is a precise cancellation of the terms proportional to $\psi^{\mu\nu}$ in \eqref{op11}. The terms proportional to the identity, on the other hand, give rise to a contribution proportional to
\begin{equation}
\text{Tr}\left[A_\mu A_\mu(A_\nu B_\nu-B_\nu A_\nu)\right]\,.
\end{equation}
While this is in general non-zero, it vanishes for the $SU(2)$ gauge group\footnote{More generally, these terms are absent if one uses a symmetric OPE as in \cite{Jakub}. These two prescriptions are related by a field redefinition.}.

Concerning the last line of \eqref{P02b}, the first term in \eqref{eq:p0mu2} can be treated in the large Hilbert space,
\begin{equation}
P_0X \xi m_2(V,V)=P_0 \xi Q \xi m_2(V,V)=\xi P_0 X m_2(V,V)=0\,,
\end{equation}
where we have used the fact that $V$ is on-shell and the last step was proven in subsection \ref{sub:exact_marginality}. The second term in \eqref{eq:p0mu2}, on the other hand, does not contain any zero mode of $\xi$. This means that the zero mode has to come from the first term in $P_0[m_2( V,\xi e_i)-m_2(\xi e_i, V)]$ (see the third line of \eqref{op11}); however, this would give rise, in the matter sector, to the one-point function $\langle\psi^{\rho\sigma}\rangle$, which is zero. Therefore the last line in \eqref{P02b} vanishes as well; this concludes the proof that $A=0$.

\subsection{Evaluation of B}
Let us now analyze the terms involving the propagator $Q^{-1}$, that is
 \begin{equation}\label{redws}
B=-2\langle X\circ X\,e_i, m_2( Q^{-1}m_2(\Psi_0,\Psi_0),\Psi_0) + m_2(\Psi_0,Q^{-1} m_2(\Psi_0, \Psi_0)\rangle\,.
\end{equation}
The field $e_i$ is of the form $e_i=c\widetilde{\mathbb{V}}_{1/2}e^{-\phi}$, where $\widetilde{\mathbb{V}}_{1/2}$ is a matter primary operator of conformal dimension $1/2$; using the picture changing we get
\begin{equation}\label{eq:XV}
Xe_i=-c\widetilde{\mathbb{V}}_1+\frac{1}{4}\gamma\widetilde{\mathbb{V}}_{1/2}\,;
\end{equation}
we now apply another picture changing operator. We consider only terms with a $c$-ghost in the final result, since they are the only ones contributing to correlation functions. For the first term in \eqref{eq:XV} only $Q_1(-\xi c\widetilde{\mathbb{V}}_1)$ maintains the $c$-ghost; for the second term we get a contribution from $Q_0(:\xi\eta:e^{\phi}\widetilde{\mathbb{V}}_{1/2})$, due to the fact that $:\xi\eta:$ is not a primary field (see appendix \ref{app:OPE} for details). We thus have, up to terms that do not contribute to the correlators,
\begin{equation}\label{eq:111}
\begin{split}
X\circ X e_i&=Q_1(-\xi c \widetilde{\mathbb{V}}_1)+Q_0(:\xi\eta:e^{\phi}\widetilde{\mathbb{V}}_{1/2})+\dots=\\
&=\oint \dfrac{dz}{2\pi i}(\eta e^\phi T_F)(z)(\xi c \widetilde{\mathbb{V}}_1)(0)+\oint \dfrac{dz}{2\pi i}(c T)(z)\left(\frac{1}{4}:\xi\eta:e^{\phi}\widetilde{\mathbb{V}}_{1/2}\right)(0)+\dots
\end{split}
\end{equation}
For the explicit calculation we notice that the supercurrent satisfies
\begin{equation}
\begin{split}\label{eq:VVT_F}
&T_F(z)\widetilde{\mathbb{V}}_{1/2}(0)=\dfrac{1}{z}\widetilde{\mathbb{V}}_1(0)+\dots\,,\\
&T_F(z)\widetilde{\mathbb{V}}_{1}(0)=\dfrac{1}{4z^2}\widetilde{\mathbb{V}}_{1/2}(z)+\mathcal{O}(z^0)\,.
\end{split}
\end{equation}
The OPE relations \eqref{eq:VVT_F} imply that the supercurrent can always be written as normal ordered product of the spacetime fermion and boson appearing in $\widetilde{\mathbb{V}}_{1/2}$ and $\widetilde{\mathbb{V}}_1$ respectively. In particular this is obvious for the gluon vertex operator, for which the spacetime fermion and boson are proportional to $\psi_\mu$ and $i\partial X^\mu$, but it is also true in the case of boundary changing operators, since we can write
\begin{equation}
T_F\propto \psi_\mu\partial X^\mu=\dfrac{1}{\sqrt{2}}:\Delta S^{\dot{\alpha}}(\bar{\sigma}_\mu)_{\dot{\alpha}\beta}S^{\beta}\tau^\mu: =\dfrac{1}{\sqrt{2}} :\bar\Delta S^{\dot{\alpha}}(\bar{\sigma}_\mu)_{\dot{\alpha}\beta}S^{\beta}\bar{\tau}^\mu:\,.
\end{equation}
Therefore \eqref{eq:111} becomes
\begin{equation}\label{eq:XX}
\begin{split}
X\circ Xe_i=&c(0)\oint \dfrac{dz}{2\pi i} e^{\phi}(z)\widetilde{\mathbb{V}}_{1/2}(z)\left(\dfrac{1}{z}+:\eta\xi:+z:\partial\eta\,\xi:+\dots\right)\left( \dfrac{1}{4z^2}+:\widetilde{\mathbb{V}}_1\widetilde{\mathbb{V}}_1:+\dots\right)+\\
&+\oint \dfrac{dz}{2\pi i} \dfrac{1}{4}c(z)\left(-\dfrac{e^{\phi}\widetilde{\mathbb{V}}_{1/2}}{z^3}+\dfrac{\partial(:\xi\eta:e^{\phi}\widetilde{\mathbb{V}}_{1/2})}{z}\right)+\dots=\\
&=\dfrac{1}{8}c\partial^2\left(e^{\phi}\widetilde{\mathbb{V}}_{1/2}\right)-\dfrac{1}{8}(\partial^2c)e^{\phi}\widetilde{\mathbb{V}}_{1/2}+ce^{\phi}\widetilde{\mathbb{V}}_{1/2}:\widetilde{\mathbb{V}}_1\widetilde{\mathbb{V}}_1:+\dfrac{1}{4}c:\partial\xi\,\eta:e^{\phi}\widetilde{\mathbb{V}}_{1/2}+\dots\,,
\end{split}
\end{equation}
where the $1/z^3$ term comes from the anomalous OPE \eqref{eq:Txieta} between the energy momentum tensor and $:\xi\eta:$ and $\dots$ indicates terms without a $c$-ghost. We notice that $ce^{\phi}\widetilde{\mathbb{V}}_{1/2}:\widetilde{\mathbb{V}}_1\widetilde{\mathbb{V}}_1:$ and $c:\partial\xi\,\eta:e^{\phi}\widetilde{\mathbb{V}}_{1/2}$ are not primary, since the OPE with the energy-momentum tensor gives
\begin{equation}
\begin{gathered}
T(z)\,\,ce^{\phi}\widetilde{\mathbb{V}}_{1/2}:\widetilde{\mathbb{V}}_1\widetilde{\mathbb{V}}_1:(0)=\dfrac{1}{4z^4}ce^{\phi}\widetilde{\mathbb{V}}_{1/2}+\dots \\
T(z)\,\,c:\partial\xi\,\eta:e^{\phi}\widetilde{\mathbb{V}}_{1/2}(0)=-\dfrac{1}{z^4}ce^{\phi}\widetilde{\mathbb{V}}_{1/2}+\dots 
\end{gathered}
\end{equation}
From these equations, however, we can see that the combination
\begin{equation}\label{VV}
ce^{\phi}\widetilde{\mathbb{V}}_{1/2}:\widetilde{\mathbb{V}}_1\widetilde{\mathbb{V}}_1:+\dfrac{1}{4}c:\partial\xi\,\eta:e^{\phi}\widetilde{\mathbb{V}}_{1/2}
\end{equation}
is a primary field, and thus behaves regularly inside the BPZ product \eqref{redws}. 

In the absence of twist field insertions these two terms will not contribute, since they give rise to one-point functions of normal ordered products. In particular, the term proportional to $:\widetilde{\mathbb{V}}_1\widetilde{\mathbb{V}}_1:$ contributes, in the matter sector, a correlator of the form 
\begin{equation}\label{VVn}
\begin{split}
&\langle \widetilde{\mathbb{V}}_{1/2}:\widetilde{\mathbb{V}}_1\widetilde{\mathbb{V}}_1:(z_1) \mathbb{V}_{1/2}(z_2)\mathbb{V}_{1/2}(z_3)\mathbb{V}_{1/2}(z_4)\rangle=\\
&=\langle :\widetilde{\mathbb{V}}_1\widetilde{\mathbb{V}}_1:(z_1)\rangle\langle 
\widetilde{\mathbb{V}}_{1/2}(z_1)\mathbb{V}_{1/2}(z_2)\mathbb{V}_{1/2}(z_3)\mathbb{V}_{1/2}(z_4)\rangle=0\,,
\end{split}
\end{equation}
where the first factor is evaluated in the un-twisted vacuum. 

We then rewrite the two remaining terms in \eqref{eq:XX} as
\begin{equation}
X\circ X\, e_i=\dfrac{1}{8}c\partial^2\left(e^{\phi}\widetilde{\mathbb{V}}_{1/2}\right)-\dfrac{1}{8}\partial^2 c(e^{\phi}\widetilde{\mathbb{V}}_{1/2})+\dots=\dfrac{1}{8}Q\left(\partial(e^{\phi}\widetilde{\mathbb{V}}_{1/2})\right)=:Q\Phi+\dots\,,
\end{equation}
up to terms that do not  not contribute to the correlation function. Since this is a $Q$-exact quantity we can compute the propagator term \eqref{redws}, which becomes
\begin{equation}
\begin{split}
B=&-2\langle Q\Phi, m_2( Q^{-1}m_2(\Psi_0,\Psi_0),\Psi_0) + m_2(\Psi_0,Q^{-1} m_2(\Psi_0, \Psi_0)\rangle=\\
=&-2\langle \Phi, m_2( (1-P_0)m_2(\Psi_0,\Psi_0),\Psi_0) - m_2(\Psi_0,(1-P_0) m_2(\Psi_0, \Psi_0)\rangle=\\
=&2\langle \Phi, m_2( P_0m_2(\Psi_0,\Psi_0),\Psi_0) - m_2(\Psi_0,P_0 m_2(\Psi_0, \Psi_0)\rangle\,,
\end{split}
\end{equation}
where the terms with the identity cancel, due to the associativity of the $m_2$ product. This can be written as
\begin{equation}\label{eq:propagator_fin}
\begin{split}
B&=2\langle P_0  m_2(\Psi_0,\Psi_0), m_2(\Psi_0,\Phi)-m_2(\Phi,\Psi_0)\rangle=\\
&=\dfrac{1}{4}\langle P_0  m_2(\Psi_0,\Psi_0), m_2(\Psi_0,\partial(e^{\phi}\widetilde{\mathbb{V}}_{1/2}))-m_2(\partial(e^{\phi}\widetilde{\mathbb{V}}_{1/2}),\Psi_0)\rangle\,.
\end{split}
\end{equation}
The operator $\partial(e^{\phi}\widetilde{\mathbb{V}}_{1/2})$ is not primary, therefore there are anomalous contributions analogous to the ones appearing in \eqref{P02b}. We refer to appendix \ref{app:anomaly} for the proof that all anomalies cancel. In the meantime we proceed as if all vertex operators were primaries, so that the product $m_2$ can be evaluated simply as the OPE. We have already computed $P_0m_2(\Psi_0,\Psi_0)$; in fact \eqref{eq:p0m2_a} gives
\begin{equation}
\left(\dfrac{\rho}{\sqrt{\alpha'}}\right)^2P_0m_2(\Psi_0,\Psi_0)=P_0m_2(V,V)=\dfrac{1}{2}\dfrac{g^2_{YM}}{\alpha'} c\partial c e^{-2\phi}\psi^{\mu\nu}M_{\mu\nu}\,,
\end{equation}
with $M_{\mu\nu}$ given in \eqref{eq:Mmunu}. On the other hand we have
\begin{equation}
\begin{split}
&P_0 \left[ m_2(V,\,\partial(e^{\phi}\widetilde{\mathbb{V}}_{1/2}))-m_2(\partial(e^{\phi}\widetilde{\mathbb{V}}_{1/2}),V)\right]=P_0 \left[V,\,\partial(e^{\phi}\widetilde{\mathbb{V}}_{1/2}) \right]_{m_2}\\
&=\lim_{z\rightarrow w}\left(\partial_w[c\mathbb{V}_{1/2}e^{-\phi}(z)e^{\phi}\widetilde{\mathbb{V}}_{1/2}(w)]-\partial_z[e^{\phi}\widetilde{\mathbb{V}}_{1/2}(z)c\mathbb{V}_{1/2}e^{-\phi}(w)] \right)=-\dfrac{g_{YM}^2}{\alpha'}c\psi^{\rho\sigma}N_{\rho\sigma}\,,
\end{split}
\end{equation}
with $N_{\rho\sigma}$ as in \eqref{eq:Nrhosigma}. Putting all together we get
\begin{equation}\label{ptg}
\left(\dfrac{\rho}{\sqrt{\alpha'}}\right)^3 B=\dfrac{1}{4}\text{Tr}\left[M_{\mu\nu}N^{\mu\nu}\right]\,,
\end{equation}
or, explicitly, assuming the ADHM constraints,
\begin{equation}\label{eq:obstr1}
\dfrac{g^4_{YM}}{8\alpha'^2}\text{Tr}\left[\left([A_\mu,A_\nu]+\dfrac{1}{2}w_{\dot{\alpha}}(\bar{\sigma}_{\mu\nu})^{\dot{\alpha}\dot{\beta}}\bar{w}_{\dot{\beta}}\right)\left([A^\mu,B^\nu]+\dfrac{1}{4}w_{\dot{\gamma}}(\bar{\sigma}^{\mu\nu})^{\dot{\gamma}\dot{\delta}}\bar{v}_{\dot{\delta}}+\dfrac{1}{4}v_{\dot{\gamma}}(\bar{\sigma}^{\mu\nu})^{\dot{\gamma}\dot{\delta}}\bar{w}_{\dot{\delta}}\right)\right]\,.
\end{equation}
In the absence of twist fields ($w_{\dot{\alpha}}=0$) this gives the correct equation of motion for a zero-momentum gluon field, in agreement with the 4-gluon vertex in Yang-Mills theory. For non-vanishing $w_{\dot{\alpha}}$, while there is a choice, as we will see later, of a zero-momentum gluon such that the anti-self-dual part of the commutator $[A_\mu,A_\nu]$ cancels the combination $\frac{1}{2}w_{\dot{\alpha}}(\bar{\sigma}_{\mu\nu})^{\dot{\alpha}\dot{\beta}}\bar{w}_{\dot{\beta}}$,  that still leaves us with the self-dual part of $[A_\mu,A_\nu]$ so that full matrix 
\begin{equation}\label{eq:AA_ww}
[A_\mu,A_\nu]+\dfrac{1}{2}w_{\dot{\alpha}}(\bar{\sigma}_{\mu\nu})^{\dot{\alpha}\dot{\beta}}\bar{w}_{\dot{\beta}}
\end{equation}
does not vanish all together, indicating an obstruction to the blow-up mode at this order. The loop-hole in this argument\footnote{This was pointed out to us by Jakub Vosmera} is that the first term in (\ref{VV}), being normal ordered w.r.t. the untwisted vacuum, may still be give a non-vanishing contribution in the twisted vacuum. It turns out that the contribution form this term is rather cumbersome to evaluate explicitly due to the presence of branch-cuts in the integrand. This difficulty can be circumvented by evaluating (\ref{redws}) in a different manner, making use of the fact that the world-sheet CFT has an $SO(4)$-invariance acting exclusively on the world-sheet fermions  $\psi^\mu$, $\mu=1,\cdots,4$ (e.g. \cite{Sen:2015uoa}) and on the spin fields. As advocated in \cite{Maccaferri:2018vwo,Maccaferri:2019ogq}, but with a slight difference due to the opposite choice of chirality for the twisted vertex operators, a convenient basis is 
\begin{align}
    \psi_1^{\pm}=\frac{1}{\sqrt{2}}\left(\psi^{1}\pm i\psi^{2}\right)\quad,\quad \psi_2^{\pm}=\frac{1}{\sqrt{2}}\left(\psi^{4}\pm i\psi^{3}\right)
\end{align}
in which only a $U(2)$ invariance is manifest. As a consequence of the $SO(4)$-invariance just described the $U(1)$-charge
\begin{align}
    J=-\frac{1}{2\pi i}\oint \sum_{i=1}^2:\psi_i^+\psi_i^-:dz=\frac{i}{2\pi i}\oint \left(\psi^{12}-\psi^{34}\right)dz
\end{align}
is conserved, with 
\begin{align}
    [J,\psi_i^+]= \psi_i^+\quad\text{and}\quad [J,\psi_i^-]= -\psi_i^- \qquad (i=1,2)\,,
\end{align}
while the spin fields have $U(1)$-charge
\begin{align}
    [J,S^{\dot{1}}]= S^{\dot{1}},\quad [J,S^{\dot{2}}]= -S^{\dot{2}} \quad\text{and}\quad [J,S^{\alpha}]= 0.
\end{align}
With our choice of chirality for the vertex operators only the spin fields with non-vanishing $U(1)$-eigenvalues will enter in the fields $\Psi_0$ and $e_i$.
Consequently, $\Psi_0$ decomposes into eigenstates of the $U(1)$-charge, i.e. $\Psi_0\mapsto \Psi_0^+ +\Psi_0^-$, in particular
\begin{equation}
    \dfrac{\rho}{\sqrt{\alpha'}}\Psi_0=V=V^+ +V^-=c{\mathbb{V}}_{1/2}^+e^{-\phi} +c\mathbb{V}_{1/2}^-e^{-\phi}\,.
\end{equation} 
An analogous decomposition holds for $e_i$, while 
\begin{equation}\label{eq:XVJ}
\dfrac{\rho}{\sqrt{\alpha'}}X\Psi_0=X V=-c{\mathbb{V}}_1+\frac{1}{4}\gamma{\mathbb{V}}_{1/2}^+ +\frac{1}{4}\gamma\mathbb{V}_{1/2}^-\,;
\end{equation}
(and analogously for $Xe_i$), where $\mathbb{V}_1$ is uncharged ($[J,{\mathbb{V}}_1]=0$), both for the twisted and untwisted sector. Upon substitution of this decomposition into (\ref{redws}) we get 
 \begin{equation}
     \begin{split}\label{redwsJ}
\langle X\circ X\,e_i, m_2( Q^{-1}m_2(\Psi_0,\Psi_0),\xi_0\Psi_0)\rangle=&+\langle X\circ X\,e_i, m_2( Q^{-1}m_2(\Psi_0^+,\Psi_0^+),\xi_0\Psi_0^-)\rangle_L\\
&+\langle X\circ X\,e_i, m_2( Q^{-1}m_2(\Psi_0^-,\Psi_0^-),\xi_0\Psi_0^+) \rangle_L\\
&-\langle X\circ X\,e_i, m_2( Q^{-1}m_2(\Psi_0^+,\xi_0\Psi_0^-),\Psi_0^+) \rangle_L\\
&+\langle X\circ X\,e_i, m_2( Q^{-1}m_2(\xi_0\Psi_0^+,\Psi_0^-),\Psi_0^-) \rangle_L\\
&+\langle X\circ X\,e_i, m_2( Q^{-1}m_2(\xi_0\Psi_0^-,\Psi_0^+),\Psi_0^+)\rangle_L\\
&-\langle X\circ X\,e_i, m_2( Q^{-1}m_2(\Psi_0^-,\xi_0\Psi_0^+),\Psi_0^-)\rangle_L
\end{split}
\end{equation}
and analogously for the second term in (\ref{redws}). Here we have used the conservation of $J$ and that, while  the $J$-charge of $X\circ X\,e_i$ can take all values form -3 to 3, in order to saturate the ghost zero-modes only the he $J$-charge $\pm 1$ part of $X\circ X\,e_i$ can contribute to the correlator. In addition the r.h.s. of (\ref{redwsJ}) is expressed in the large Hilbert space. The position of the $\xi$-zero mode is correlated with relative sign of each term. Next we write $ X\circ X\,e_i=Q \xi\circ X\,e_i$ and bring the BRST charge $Q$ to the other side through BPZ-conjugation. The only contribution comes from the commutator $\{Q, Q^{-1}\}$ since, whenever $Q$ hits a $\xi$, the $J$-charge does not add up to zero or the ghost zero-modes are not saturated. Adding in the second term on the r.h.s. of (\ref{redws}) we are left with
\begin{equation}
     \begin{split}\label{redwsJx}
-\frac12 B=&\langle \xi\circ X\,e_i, m_2( P_0m_2(\Psi_0^+,\Psi_0^+),\xi \Psi_0^-)- m_2( \Psi_0^+,P_0m_2(\Psi_0^+,\xi \Psi_0^-)\rangle_L\\
&+\langle \xi\circ X\,e_i, m_2( P_0m_2(\Psi_0^-,\Psi_0^-),\xi \Psi_0^+) -m_2(\Psi_0^-,P_0m_2(\Psi_0^-,\xi \Psi_0^+) \rangle_L\\
&-\langle \xi\circ X\,e_i, m_2( P_0m_2(\Psi_0^+,\xi \Psi_0^-),\Psi_0^+) + m_2( \Psi_0^+,P_0m_2(\xi \Psi_0^-,\Psi_0^+)\rangle_L\\
&+\langle \xi\circ X\,e_i, m_2( P_0m_2(\xi \Psi_0^+,\Psi_0^-),\Psi_0^-)-m_2( \xi \Psi_0^+,P_0m_2(\Psi_0^-,\Psi_0^-) \rangle_L\\
&+\langle \xi\circ X\,e_i, m_2( P_0m_2(\xi \Psi_0^-,\Psi_0^+),\Psi_0^+)- m_2( \xi \Psi_0^-,P_0m_2(\Psi_0^+,\Psi_0^+)\rangle_L\\
&-\langle \xi\circ X\,e_i, m_2( P_0m_2(\Psi_0^-,\xi \Psi_0^+),\Psi_0^-)+ m_2( \Psi_0^-,P_0m_2(\xi \Psi_0^+,\Psi_0^-)\rangle_L\,,
\end{split}
\end{equation}
where we have used the associativity of $m_2$. With the help of the cyclic property (\ref{cyclic}) of $m_2$ this can be recast into
\begin{align}\label{redwsJxc}
-\frac12 B=&\langle P_0m_2(\Psi_0^+,\Psi_0^+),[\xi \Psi_0^-,\xi X e_i]_{m_2}\rangle_L+\langle P_0m_2(\Psi_0^-,\Psi_0^-),[\xi \Psi_0^+,\xi X e_i]_{m_2} \rangle_L\nonumber\\
&-\langle P_0[\Psi_0^+,\xi \Psi_0^-]_{m_2},[\Psi_0^+,\xi X e_i]_{m_2}\rangle_L+\langle P_0[\xi \Psi_0^+,\Psi_0^-]_{m_2},[\Psi_0^-,\xi  X e_i]_{m_2} \rangle_L\,.
\end{align}
The four contributions to the r.h.s. of (\ref{redwsJxc}) can be read-off from eqns. (\ref{eq:p0m2_a}-\ref{eq:Nrhosigma}). Explicitly we have
\begin{equation}
\begin{split}
    &\left(\dfrac{\rho}{\sqrt{\alpha'}}\right)^2 P_0m_2(\Psi_0^\pm,\Psi_0^\pm)= P_0m_2(V^\pm,V^\pm)=-\dfrac{1}{4}\dfrac{g_{YM}^2}{\alpha'}c\partial c e^{-2\phi}\bar{\eta}^{\mu\nu}_\mp M_{\mu\nu}\psi_{12}^{\pm\pm}\,,\\
    & \left(\dfrac{\rho}{\sqrt{\alpha'}}\right)P_0[\xi\Psi_0^\pm,\xi X e_i^\pm]_{m_2}=P_0[\xi V^\pm,\xi X e_i^\pm]_{m_2}=\dfrac{1}{8}\dfrac{g_{YM}^2}{\alpha'}\xi_0 c \bar{\eta}^{\rho\sigma}_\mp N_{\rho\sigma}\psi_{12}^{\pm\pm}+\dots\,,\\
    &\left(\dfrac{\rho}{\sqrt{\alpha'}}\right)^2P_0[\Psi_0^\pm,\xi \Psi_0^\mp]_{m_2}=P_0[V^\pm,\xi V^\mp]_{m_2}=\pm\dfrac{i}{4}\dfrac{g_{YM}^2}{\alpha'} \partial\xi c\partial c e^{-2\phi}\bar{\eta}^{\mu\nu}_3 M_{\mu\nu}+\dots\,,\\
    &\left(\dfrac{\rho}{\sqrt{\alpha'}}\right)P_0[\Psi_0^\pm,\xi X e_i^\mp]_{m_2}=P_0[V^\pm,\xi X e_i^\mp]_{m_2}=\mp\dfrac{i}{8} c:\xi\eta:\bar{\eta}^{\rho\sigma}_3 M_{\rho\sigma}\,,
\end{split}
\end{equation}
where $\bar{\eta}^{\mu\nu}_{\pm}=\bar{\eta}^{\mu\nu}_1\pm i \bar{\eta}^{\mu\nu}_2$ are defined in terms of the 't Hooft symbols, and $M_{\mu\nu}$ and $N_{\rho\sigma}$ are matrices defined above. The $\dots$ denote terms that vanish upon insertion in the inner product in (\ref{redwsJxc}).   Putting all together we end up with
\begin{equation}
\left(\dfrac{\rho}{\sqrt{\alpha'}}\right)^3 B=\dfrac{1}{8}\text{Tr}\left[M^a N^a \right]\,,
\end{equation}
where the matrices $M^a$ and $N^a$ are as
\begin{equation}
\begin{gathered}
M^a=\bar{\eta}^a_{\mu\nu}\left([A^\mu,A^\nu]+\dfrac{1}{2}w_{\dot{\alpha}}(\bar{\sigma}^{\mu\nu})^{\dot{\alpha}\dot{\beta}}\bar{w}_{\dot{\beta}}\right)\,,\\
N^a=\bar{\eta}^a_{\mu\nu}\left([A^\mu,B^\nu]+\dfrac{1}{4}w_{\dot{\gamma}}(\bar{\sigma}^{\mu\nu})^{\dot{\gamma}\dot{\delta}}\bar{v}_{\dot{\delta}}+\dfrac{1}{4}v_{\dot{\gamma}}(\bar{\sigma}^{\mu\nu})^{\dot{\gamma}\dot{\delta}}\bar{w}_{\dot{\delta}}\right)\,.
\end{gathered}
\end{equation}
Notice that this reproduces (\ref{ptg}), however, with the important difference that $\text{Tr}\left(2M_{\mu\nu}N^{\mu\nu}\right)$ is replaced by $\text{Tr}\left(M^a N^a\right)$. 
This can be seen clearly if we rewrite, assuming the ADHM constraints,
\begin{equation}\label{eq:obstr2}
\begin{split}
\text{Tr}(M_{\mu\nu}N^{\mu\nu})=&\dfrac{1}{2}\text{Tr}\left(M^a N^a\right)-\text{Tr}\left[[A_\mu,A_\nu] \left(\dfrac{1}{4}w_{\dot{\gamma}}(\bar{\sigma}^{\mu\nu})^{\dot{\gamma}\dot{\delta}}\bar{v}_{\dot{\delta}}+\dfrac{1}{4}v_{\dot{\gamma}}(\bar{\sigma}^{\mu\nu})^{\dot{\gamma}\dot{\delta}}\bar{w}_{\dot{\delta}}\right)\right]+\\
&-\text{Tr}\left([A_\mu,B_\nu]\,\dfrac{1}{2}w_{\dot{\alpha}}(\bar{\sigma}_{\mu\nu})^{\dot{\alpha}\dot{\beta}}\bar{w}_{\dot{\beta}} \right)\,.
\end{split}
\end{equation}
This means that the contributions coming from \eqref{VVn} in the twisted sector have the effect of exactly cancelling all the terms in $\text{Tr}[M_{\mu\nu}N^{\mu\nu}]$ that are not anti-self-dual\footnote{In the first version of this paper this difference was missed because (\ref{VVn}) was assumed to hold also in the presence of twist fields.}, leaving only terms proportional to $\text{Tr}\left(M^a N^a\right)$.
It is then possible, in agreement with \cite{Maccaferri:2019ogq, Jakub}, to set $M^a$ to zero assuming the ADHM constraints \eqref{eq:ADHM}
\begin{equation}
\bar{\eta}_a^{\mu\nu}\left([a_\mu,a_\nu]+\frac{1}{2}\bar{w}_{\dot{\alpha}}(\bar{\sigma}_{\mu\nu})^{\dot{\alpha}\dot{\beta}}w_{\dot{\beta}}\right)=0\,,\\
\end{equation}
and with a suitable choice of the matrices $A_\mu$, that is
\begin{equation}
    A_\mu=\frac{\rho}{\sqrt{2}}\sigma_\mu=\frac{\rho}{\sqrt{2}}(\unit, -i\vec\tau )\,.
\end{equation} 
As discussed in section \ref{subs:profile}, this zero momentum gluon contributes to the instanton profile at order $\rho$ (see \eqref{eq:full_profile}) but not at order $\rho^2$. Furthermore, it is in principle possible to compute all contributions to the instanton profile at order $\mathcal{O}(\rho^3)$, inverting \eqref{eq:psi2}. The explicit calculation is, however, quite involved.

\section{Conclusions}
The motivation for this work was to better understand bound states of D-branes in superstring theory. In particular, we focused on the D(-1)-D3 brane system, since the corresponding field theory in four dimensions is the well-known $\mathcal{N}=4$ SYM theory which admits pointlike instantons as singular, non-perturbative configurations, which are recovered from the D(-1)-D3 brane bound state in the field theory limit. In this paper we studied the possibility to extend this  connection to finite size (not pointlike) D(-1) branes inside a D3 background, constructing them as marginal deformation of the worldsheet theory of pointlike D(-1) branes. 

The standard worldsheet approach can not be applied here for two reasons. First, the computation of the instanton profile is a off-shell problem in string theory; second, there are subtleties with the integration over odd moduli in super moduli space which are not captured by the worldsheet description. In the present paper we deal with this problem by working with the  $A_\infty$ SFT.

After reviewing the derivation of the instanton profile in the small size ($\rho/\sqrt{\alpha'}\ll 1$) and far distance ($\sqrt{\alpha'}/|x-x_0|\ll 1$) limit, we extended this result to all orders in $\sqrt{\alpha'}/|x-x_0|$. We also studied the deformation corresponding to the blow up mode of the size of a D(-1) brane inside a D3 background. This deformation was found to be marginal at second order in size, insensitive of the ADHM constraints, while at third order in addition to the ADHM constraints an addition zero-momentum gluon is required for marginality. 

An interesting question to explore is whether it is possible to find more generic hermitian string fields that are solutions to the equations of motion not satisfying the ADHM constraints. We were not able to find any such solutions but cannot exclude them on general grounds at this point. 

While we considered the specific case of the D(-1)-D3 brane bound state in this paper, our approach applies equally well to generic D$p$-D$(p+4)$ brane bound states. Another interesting extension concerns the blow up of orbifold singularities in closed super string field theory \cite{Dixon:1985jw,Baba:2009ns,Erler:2014eba,Sen:2015uaa}. We hope to come back to this question in the near future.

\newpage
\noindent{\bf Acknowledgements:}\\
We would like to thank Carlo Maccaferri, Igor Pesando, Martin Schnabl, Ashoke Sen 
for constructive comments and especially Jakub Vo\v{s}mera for pointing out important corrections to the first version of this paper. In addition we would like to thank Enrico Andriolo, Sebastian Konopka, Alberto Merlano and Tom\'{a}\v{s} Proch\'{a}zka for inspiring  discussions. We thank the Galileo Galilei Institute for Theoretical Physics and INFN for hospitality and partial support during the workshop ``String Theory from a worldsheet perspective". This work has been supported by the Excellence Cluster ``Origins: From the Origin of the Universe to the First Building Blocks of Life''.

\newpage
\appendix
\section{Notation and Conventions}\label{app:conventions}

\subsection*{Notation for indices}
In this work we use many indices with different meanings. The most used ones are the following:
\begin{itemize}
\item $d=4$ vector indices: $\mu,\nu=0,\dots,3$;
\item $d=6$ vector indices: $a,b=4,\dots,9$;
\item Chiral and anti-chiral spinor indices in $d=4$: $\alpha$ and $\dot{\alpha}$;
\item Spinor indices in $d=6$: $^A$ and $_A$ in the fundamental and anti-fundamental of $SU(4)\simeq SO(6)$;
\item D3 indices: $u,v=1,\dots,N$;
\item D(-1) indices: $i,j=1,\dots,k$;
\item $SU(2)$ colour indices: $c,d=1,2,3$.
\end{itemize}

\subsection*{d=4 Clifford algebra and spinors}
In $d=4$ we can either deal with the Euclidean (SO(4)) or Minkowskian (SO(1,3)) Lorentz group; its Clifford algebra is defined by $\{\gamma^{\mu},\gamma^{\nu}\}=2\eta^{\mu\nu}\unit$, where the metric $\eta$ has signature $(+,+,+,+)$ or $(-,+,+,+)$ respectively. 
Let us consider the Pauli matrices $\tau^c$
\begin{equation}
\tau^1=\begin{pmatrix} 0 & 1 \\ 1 & 0 \end{pmatrix}\,, \qquad 
\tau^2=\begin{pmatrix} 0 & -i \\ i & 0 \end{pmatrix}\,, \qquad
\tau^3=\begin{pmatrix} 1 & 0 \\ 0 & -1 \end{pmatrix}\,.
\end{equation}
Gamma matrices in four dimensions can be expressed in terms of the matrices $(\sigma^\mu)_{\alpha\dot{\beta}}$ and $(\bar{\sigma}^\mu)^{\dot{\alpha}\beta}$ in the following way:
\begin{equation}
\gamma^\mu= \begin{pmatrix}
0 & \sigma^\mu \\
\bar{\sigma}^\mu & 0
\end{pmatrix}\,,
\end{equation}
where $\sigma^\mu$ and $\bar{\sigma}^\mu$ are defined in terms of the Pauli matrices, but in a different way for Euclidean and Minkowski space:
\begin{equation}
\begin{split}
&\sigma^\mu=(\unit,-i\vec{\tau}) \quad \mbox{and} \quad \bar{\sigma}^\mu=(\unit,i\vec{\tau}) \qquad\mbox{(Euclidean)} \\
&\sigma^\mu=(\unit,\vec{\tau}) \quad \mbox{and} \quad \bar{\sigma}^\mu=(-\unit,\vec{\tau}) \qquad\,\,\,\,\mbox{(Minkowski)}
\end{split}
\end{equation}
They satisfy the appropriate Clifford algebra $\sigma^\mu\bar{\sigma}^\nu+\sigma^\nu\bar{\sigma}^\mu=2\eta^{\mu\nu}\unit$.

It is convenient to divide every Dirac spinor into its two Weyl components as follows:
\begin{equation}
\psi=\begin{pmatrix}
\psi_\alpha \\ \psi^{\dot{\alpha}}
\end{pmatrix}\,.
\end{equation}
We raise and lower spinor indices contracting always with the second index of the antisymmetric $\varepsilon$ tensor:
\begin{equation}
\psi^{\alpha}=\varepsilon^{\alpha\beta}\psi_\beta \quad , \quad \psi_{\dot{\alpha}}=\varepsilon_{\dot{\alpha}\dot{\beta}}\psi^{\dot{\beta}}\,,
\end{equation}
with $\varepsilon^{12}=\varepsilon_{12}=-\varepsilon^{\dot{1}\dot{2}}=-\varepsilon_{\dot{1}\dot{2}}=1$. Therefore, we have also
\begin{equation}
\psi_{\alpha}=\psi^\beta\varepsilon_{\beta\alpha} \quad , \quad \psi^{\dot{\alpha}}=\psi_{\dot{\beta}}\varepsilon^{\dot{\beta}\dot{\alpha}}\,.
\end{equation}
Depending on the metric $\eta$, the $\sigma$ matrices behave differently under complex conjugation. In Euclidean space one has
\begin{equation}
(\sigma^\mu)_{\alpha\dot{\beta}}^*=-(\sigma^\mu)^{\alpha\dot{\beta}}\qquad \mbox{and} \qquad (\bar{\sigma}^\mu)^{\dot{\alpha}\beta*}=-(\bar{\sigma}^\mu)_{\dot{\alpha}\beta}\,,
\end{equation}
while in Minkowski space one type of index gets changed into the other:
\begin{equation}
(\sigma^\mu)_{\alpha\dot{\beta}}^*=(\sigma^\mu)_{\beta\dot{\alpha}}\,.
\end{equation}
Both in Minkowskian and in Euclidean case we have the following important relation:
\begin{equation}
(\sigma^\mu)_{\alpha\dot{\beta}}=(\bar{\sigma}^\mu)_{\dot{\beta}\alpha}\,.
\end{equation}

\subsection*{Euclidean d=4 Clifford algebra and 't Hooft symbols}
In the following we focus only on the Euclidean case, because it is the one we are interested in when dealing with instantons. The SO(4) generators are defined in terms of $\sigma$ matrices in the following way:
\begin{equation}
\sigma_{\mu\nu}=\dfrac{1}{2}(\sigma_\mu\bar{\sigma}_\nu-\sigma_\nu\bar{\sigma}_\mu) \quad ,\quad \bar{\sigma}_{\mu\nu}=\dfrac{1}{2}(\bar{\sigma}_\mu\sigma_\nu-\bar{\sigma}_\nu\sigma_\mu) \,;
\end{equation}
These matrices satisfy self-duality or anti-self-duality conditions respectively, in particular:
\begin{equation}
\sigma_{\mu\nu}=\dfrac{1}{2}\varepsilon_{\mu\nu\rho\sigma}\sigma_{\rho\sigma}\quad ,\quad \bar{\sigma}_{\mu\nu}=-\dfrac{1}{2}\varepsilon_{\mu\nu\rho\sigma}\bar{\sigma}_{\rho\sigma}\,.
\end{equation}
The mapping between a self-dual (or anti-self-dual) SO(4) tensor into the corresponding adjoint representation of SU(2) is given in terms of the 't Hooft symbols as follows:
\begin{equation}\label{eq:sigma_eta}
 (\sigma_{\mu\nu})_\alpha^{\;\;\beta}=i\eta^c_{\mu\nu}(\tau^c)_\alpha^{\;\;\beta} \quad ,\quad  (\bar{\sigma}_{\mu\nu})^{\dot{\alpha}}_{\;\;\dot{\beta}}=i\bar{\eta}^c_{\mu\nu}(\tau^c)^{\dot{\alpha}}_{\;\;\dot{\beta}}\,.
 \end{equation} 
An explicit representations of these symbols is given by:
\begin{equation}
\begin{split}
&\eta^c_{\mu\nu}=\bar{\eta}^c_{\mu\nu}=\varepsilon_{c\mu\nu}\,, \qquad \mu,\nu\in\{1,2,3\} \\
&\eta^c_{0\nu}=-\bar{\eta}^c_{0\nu}=\delta^c_\nu\,, \\
&\eta^c_{\mu\nu}=-\eta^c_{\nu\mu}\,, \\
&\bar{\eta}^c_{\mu\nu}=-\bar{\eta}^c_{\nu\mu}\,.
\end{split}
\end{equation}
Many properties of these symbols can be found in the literature. In particular the symbols $\eta^c_{\mu\nu}$ and $\bar{\eta}^c_{\mu\nu}$ are self-dual and anti-self-dual respectively.

\section{Relevant Operators and their OPE's}\label{app:OPE}
In the calculation of amplitudes in a conformal field theory it is important to know the operator product expansion (OPE) of primary fields $\mathcal{O}(z_1)\mathcal{O}(z_2)$.
First of all, let us consider the primary fields $\partial X_\mu(z)$, with conformal weight $1$. The OPE of two of them is
\begin{equation}
\partial X^\mu(z) \partial X^\nu(w)=-\dfrac{\alpha'}{2}\dfrac{\delta^{\mu\nu}}{(z-w)^2}+\dots\,,
\end{equation}
where $\dots$ indicate regular terms.
For the spinors $\psi^\mu$ the OPE is given by
\begin{equation}
\psi^\mu(z) \psi^\nu(w)=\dfrac{\delta^{\mu\nu}}{z-w}+\psi^{\mu\nu}+\dots\,.
\end{equation}

The presence of D3 branes breaks $SO(10)$ to $SO(4)\times SO(6)$ (we consider the euclidean theory); therefore the ten dimensional spin fields $S^{\mathcal{A}}$ and $S^{\dot{\mathcal{A}}}$ can be expressed in terms of the spin fields in 4 and 6 dimensions as follows:
\begin{equation}
\begin{split}
S^{\mathcal{A}} \longrightarrow (S_{\alpha}S^A, S^{\dot{\alpha}}S_A)\,, \\
S^{\dot{\mathcal{A}}} \longrightarrow (S_{\alpha}S_A, S^{\dot{\alpha}}S^A)\,,
\end{split}
\end{equation}
where $S_\alpha$ and $S^{\dot{\alpha}}$ are $SO(4)$ spin fields of even and odd chirality respectively, and $S^A$ and $S_A$ are $SO(6)$ spin fields of even and odd chirality respectively. Spin fields in $d=4$ can be bosonized with exponents:
\begin{equation}
\begin{split}
\lambda_{\alpha}=\left(\dfrac{1}{2},\dfrac{1}{2}\right) \quad \mbox{or} \quad \left(-\dfrac{1}{2},-\dfrac{1}{2}\right)\,, \\
\lambda_{\dot{\alpha}}=\left(\dfrac{1}{2},-\dfrac{1}{2}\right) \quad \mbox{or} \quad \left(-\dfrac{1}{2},\dfrac{1}{2}\right)\,.
\end{split}
\end{equation}
Their OPE contains branch cuts; explicitly we have
\begin{equation}
\begin{split}
&S^{\dot{\alpha}}(z)S_{\beta}(w)=\dfrac{1}{\sqrt{2}}(\bar{\sigma}^\mu)^{\dot{\alpha}}_{\;\;\beta}\psi_\mu(w)+\dots\,, \\
&S^{\dot{\alpha}}(z)S^{\dot{\beta}}(w)=-\dfrac{\varepsilon^{\dot{\alpha}\dot{\beta}}}{(z-w)^{1/2}}+\dfrac{1}{4}(z-w)^{1/2}(\bar{\sigma}_{\mu\nu})^{\dot{\alpha}\dot{\beta}}\psi^{\mu\nu}+\dots\,, \\
&S_{\alpha}(z)S_{\beta}(w)=\dfrac{\varepsilon_{\alpha\beta}}{(z-w)^{1/2}}-\dfrac{1}{4}(z-w)^{1/2}(\sigma_{\mu\nu})_{\alpha\beta}\psi^{\mu\nu}+\dots\,.
\end{split}
\end{equation}
All these expressions can be derived using the bosonization of the spin fields; furthermore, one can also derive the following OPE involving spinors and spin fields:
\begin{equation}\label{eq:psi_S}
\begin{split}
&\psi_\mu(z)S^{\dot{\alpha}}(w)=\dfrac{1}{\sqrt{2}}\dfrac{(\bar{\sigma}_\mu)^{\dot{\alpha}\beta}S_\beta(w)}{(z-w)^{1/2}}+\dots\,,\\
&\psi_{\mu\nu}(z)S^{\dot{\alpha}}(w)=-\dfrac{1}{2}\dfrac{(\bar{\sigma}_{\mu\nu})^{\dot{\alpha}}_{\;\;\dot{\beta}}S^{\dot{\beta}}(w)}{z-w}+\dots\,.
\end{split}
\end{equation}
From these OPE one can easily compute some three-point functions, for example:
\begin{equation}
\begin{split}
&\langle S^{\dot{\alpha}}(z_1)\psi_\mu(z_2)S_{\beta}(z_3)\rangle=\dfrac{1}{\sqrt{2}}\dfrac{(\bar{\sigma}_\mu)^{\dot{\alpha}}_{\;\;\beta}}{z_{12}^{1/2}z_{23}^{1/2}}\,,\\
&\langle S^{\dot{\alpha}}(z_1)\psi_{\mu\nu}(z_2)S^{\dot{\beta}}(z_3)\rangle=-\dfrac{1}{2}(\bar{\sigma}_{\mu\nu})^{\dot{\alpha}\dot{\beta}}\dfrac{z_{13}^{1/2}}{z_{12}z_{23}}\,,
\end{split}
\end{equation}
where we have introduced the notation $z_{ij}=z_i-z_j$. Other details on the spin fields and their bosonization can be found in \cite{Kostelecky:1986xg,Polchinski:1996na}.

Regarding the twist operators, we have to deal with non trivial OPE with the fields $\partial X^\mu$ and $e^{ik\cdot X}$. Remembering that the field $\Delta(z)$ is made of four twist operators ($\Delta(z)=\sigma^0\sigma^1\sigma^2\sigma^3(z)$), it is sufficient to know the behavior of one field $\sigma^{\mu}(z)$, which has conformal dimension $1/16$. We have the following relevant OPE, involving also the so-called excited twist field $\sigma'^{\mu}(z)$, with conformal dimension $9/16$ \cite{Mattiello:2018kue}:
\begin{equation}\label{eq:dX_s}
\begin{split}
&\sigma^{\mu}(z)\bar{\sigma}^{\nu}(w)=\dfrac{\delta^{\mu\nu}}{(z-w)^{1/8}}+\dots\,,\\
&\sqrt{\dfrac{2}{\alpha'}}i\partial X^{\mu}(z)\sigma^{\nu}(w)=\dfrac{\delta^{\mu\nu}\sigma'^{\nu}(w)}{(z-w)^{1/2}}+\dots\,,\\
&\sqrt{\dfrac{2}{\alpha'}}i\partial X^{\mu}(z)\sigma'^{\nu}(w)=\dfrac{1}{2}\dfrac{\delta^{\mu\nu}\sigma^{\nu}(w)}{(z-w)^{3/2}}+\dfrac{2\delta^{\mu\nu}\partial \sigma^{\nu}(w)}{(z-w)^{1/2}}+\dots\,,
\end{split}
\end{equation}
where we do not sum over equal indices. From these OPE one can derive the three-point function
\begin{equation}
\langle\bar{\Delta}(z_1)e^{-ik\cdot X(z_2)}\Delta(z_3)\rangle=\dfrac{e^{-ik\cdot x_0}}{(z_{13})^{1/2-\alpha'k^2/2}(4z_{12}z_{23})^{\alpha'k^2/2}}\,,
\end{equation}
where $x_0^\mu$ is the zero-mode of the field $X^{\mu}(z)$. More properties of these twist operators can be found for example in \cite{Mattiello:2018kue,Zamolodchikov:1987ae}.

In superstring theory one has also to deal with ghosts and superghosts, which are characterized by the OPE relations
\begin{equation}
\begin{split}
&b(z)c(w)= c(z)b(w)\sim\dfrac{1}{z-w}+\dots\,,\\
&c(z)c(w)= -(z-w)c\partial c(w)-\dfrac{1}{2}(z-w)^2c\partial^2c(w)+\dots\,,\\
&\beta(z)\gamma(w)\sim-\gamma(z)\beta(w)= -\dfrac{1}{z-w}+\dots\,.
\end{split}
\end{equation}
These ghosts can be bosonized in the following way
\begin{equation}\label{eq:superghosts}
\begin{split}
&b=e^{-\sigma}\,,\qquad c=e^{\sigma}\,,\\
&\beta=e^{-\phi}\partial\xi=e^{-\phi}e^{\chi}\partial\chi\,,\qquad \gamma=\eta e^{\phi}=e^{-\chi}e^{\phi}\,,
\end{split}
\end{equation}
with the following OPE relations
\begin{equation}
\begin{split}
&\sigma(z)\sigma(w)=\log(z-w)+\dots\,,\\
&\phi(z)\phi(w)= -\log (z-w)+\dots\,,\\
&\chi(z)\chi(w)= \log (z-w)+\dots\,,\\
&\xi(z)\eta(w)=\eta(z)\xi(w)\sim\dfrac{1}{z-w}+\dots\,,\\
&e^{-\phi}(z)e^{\phi}(w)\sim e^{\phi}(z)e^{-\phi}(w) =(z-w)+\dots\,,\\
&e^{-\phi}(z)e^{-\phi}(w)=\dfrac{1}{z-w}e^{-2\phi}(w)-\partial\phi e^{-2\phi}(w)+\dots\,.
\end{split}
\end{equation}
The relevant two- and three-point functions used in this work are the following ones:
\begin{equation}
\begin{split}
&\langle c(z_1)c(z_2)c(z_3)\rangle=z_{12}z_{23}z_{13}\,,\\
&\langle e^{-\phi(z_1)}e^{-\phi(z_2)}\rangle=\dfrac{1}{z_{12}}\,,\\
&\langle c\partial c e^{-2\phi}(z)c(w)\rangle=-(z-w)^2\,.
\end{split}
\end{equation}

\subsection*{Non-Primary Operators}
In this paper we have to deal with some operators that are not primary. In particular we encounter $:\xi\eta:e^{\phi}\widetilde{\mathbb{V}}_{1/2}$ and $\partial(e^{\phi}\widetilde{\mathbb{V}}_{1/2})$. The normal ordered product is defined in terms of the OPE as
\begin{equation}
:\xi\eta:(w)=\oint \dfrac{dx}{2\pi i}\dfrac{\xi(x)\eta(w)}{x-w};
\end{equation}
$\xi$ and $\eta$ are primaries, thus we can compute 
\begin{equation}\label{eq:Txieta}
T(z):\xi\eta:(0)=T(z)\oint \dfrac{dx}{2\pi i}\dfrac{\xi(x)\eta(0)}{x} =-\dfrac{1}{z^3}+\dfrac{:\xi\eta:(0)}{z^2}+\dfrac{\partial:\xi\eta:(0)}{z}+\dots
\end{equation}
The presence of a cubic pole shows that $:\xi\eta:$ is not a primary operator; from this we derive
\begin{equation}\label{eq:Txieta_bis}
T(z):\xi\eta:e^{\phi}\widetilde{\mathbb{V}}_{1/2}(0)=T(z)\oint \dfrac{dx}{2\pi i}\dfrac{\xi(x)\eta(0)}{x} =-\dfrac{e^{\phi}\widetilde{\mathbb{V}}_{1/2}(0)}{z^3}+\dfrac{\partial(:\xi\eta:e^{\phi}\widetilde{\mathbb{V}}_{1/2})(0)}{z}+\dots
\end{equation}
Similarly, for $\partial(e^{\phi}\widetilde{\mathbb{V}}_{1/2})$ we get
\begin{equation}\label{eq:Tderiv}
T(z)\partial(e^{\phi}\widetilde{\mathbb{V}}_{1/2})(0)=-2\dfrac{e^{\phi}\widetilde{\mathbb{V}}_{1/2}(0)}{z^3}+\dfrac{\partial^2(e^{\phi}\widetilde{\mathbb{V}}_{1/2})(0)}{z}+\dots
\end{equation}

\section{Details on the Calculation of the Instanton Profile}\label{app:profile}
In this appendix we discuss in detail the calculation of the instanton profile sketched in section \ref{subs:profile}. We start from 
\begin{equation}\label{eq:prof_app_1}
A_{\mu}^{c(1)}(k)=C_0 f_1'(0)^{\alpha'k^2/2}\langle V_{\bar{w}}^{(-1)u}(\infty)\mathcal{V}^{(0)uv}_{A_{\mu}}(1;-k)V_{w}^{(-1)v}(0) \rangle\,.
\end{equation}
The boundary changing operators (in picture -1) are the ones given in \eqref{eq:vertex_3-1_NS} with the rescaling \eqref{eq:rescaling}, while $\mathcal{V}_{A_\mu}$ is given (in picture 0) by \eqref{eq:gluon_vo}. We also compute the correlation function at generic positions $z_1$, $z_2$ and $z_3$, and then consider the particular case $z_1\rightarrow\infty$, $z_2=1$ and $z_3=0$. We can split the amplitude \eqref{eq:prof_app_1} in four sub-amplitudes, which are independent from each other because they contain fields belonging to different CFT's:
\begin{equation}\label{eq:firstorder1}
\begin{split}
A_{\mu}^{c(1)}(k)\sim\bar{w}_{\dot{\alpha}}^u(\tau^c)^{vu}&w_{\dot{\beta}}^v k^\nu\langle c(z_1)c(z_2)c(z_3)\rangle\langle e^{-\phi(z_1)}e^{-\phi(z_3)}\rangle\cdot \\ &\cdot\langle\bar{\Delta}(z_1)e^{-ik\cdot X}(z_2)\Delta(z_3)\rangle\langle S^{\dot{\alpha}}(z_1)\psi_{\nu}\psi_{\mu}(z_2)S^{\dot{\beta}}(z_3)\rangle\,.
\end{split}
\end{equation}
The first term in \eqref{eq:gluon_vo} has not been taken into account: its contribution would be proportional to $k_{\mu}$; anyway, the polarization $A^\mu$ of the vector is subjected to the constraint $A\cdot k=0$. Notice that all factors of $\alpha'$ (except the exponent of $f_1'(0)$) and $g_{YM}$ disappear, thanks to the rescaling \eqref{eq:rescaling}. All the correlation functions appearing in \eqref{eq:firstorder1} are well known (see appendix \ref{app:OPE}); we can thus write
\begin{equation}
\begin{split}
A_{\mu}^{c(1)}(k)\sim f_1'(0)^{\alpha'k^2/2}\bar{w}_{\dot{\alpha}}^u&(\tau^c)^{vu} w_{\dot{\beta}}^v k^\nu \, (z_{12}z_{23}z_{13}) \, \left(\dfrac{1}{z_{13}} \right)\cdot\\
\cdot&\left(\dfrac{e^{-ik\cdot x_0}}{z_{13}^{(1-\alpha'k^2)/2}(4z_{12}z_{23})^{\alpha'k^2/2}} \right)\, \left(-\dfrac{1}{2}(\bar{\sigma}_{\nu\mu})^{\dot{\alpha}\dot{\beta}}\dfrac{z_{13}^{1/2}}{z_{12}z_{23}} \right)\,,
\end{split}
\end{equation}
where the 4-vector $x_0^\mu$ denotes the position of the D(-1) brane inside the D3 brane.
Simplifying the result we are left with
\begin{equation}\label{eq:prof_momentum_generic}
A_{\mu}^{c(1)}(k)\sim\left(\dfrac{f_1'(0)z_{13}}{4z_{12}z_{23}}\right)^{\alpha'k^2/2}\dfrac{1}{2}\bar{w}_{\dot{\alpha}}^u (\bar{\sigma}_{\nu\mu})^{\dot{\alpha}}_{\;\;\dot{\beta}} w^{v\dot{\beta}}(\tau^c)^{vu} k^\nu\, e^{-ik\cdot x_0}\,.
\end{equation}
It is convenient now to use the 't Hooft symbols (see \eqref{eq:sigma_eta}); we then obtain
\begin{equation}
A_{\mu}^{c(1)}(k)\sim\left(\dfrac{f_1'(0)z_{13}}{4z_{12}z_{23}}\right)^{\alpha'k^2/2}\,\dfrac{i}{2}\bar{\eta}^d_{\nu\mu}\left(\bar{w}_{\dot{\alpha}}^u (\tau^d)^{\dot{\alpha}}_{\;\;\dot{\beta}} w^{v\dot{\beta}}\right)(\tau^c)^{vu} k^\nu\, e^{-ik\cdot x_0}\,.
\end{equation}
Using the solution to the ADHM constraint
\begin{equation}\label{eq:size}
w^{v\dot{\beta}}\bar{w}_{\dot{\alpha}}^{v}=\rho^2\delta_{\dot{\alpha}}^{\dot{\beta}}\,,
\end{equation}
we can see that the $N\times N$ matrices
\begin{equation}
(t^d)^{uv}=\dfrac{1}{\rho^2}\left(\bar{w}_{\dot{\alpha}}^u(\tau^d)^{\dot{\alpha}}_{\;\;\dot{\beta}}w^{v\dot{\beta}}\right)
\end{equation}
satisfy the relation $[t^d,t^e]=2i\epsilon^{def}t^f$. If the solution \eqref{eq:ADHM_sol} is considered, we can identify them with the Pauli matrices $(t^d)^{uv}=(\tau^d)^{uv}$. We can then conclude that (rescaling the size $\rho$ if necessary, and setting the three points to $\infty$, 1 and 0 respectively)
\begin{equation}\label{eq:instanton_1_momentum}
A_{\mu}^{c(1)}(k)=\left(\dfrac{f_1'(0)}{4}\right)^{\alpha'k^2/2}\dfrac{i\rho^2}{2}k^\nu\bar{\eta}^d_{\nu\mu}e^{-ik\cdot x_0}\mbox{Tr}(\tau^d\tau^c)=f_1'(0)^{\alpha'k^2/2}i\rho^2 k^\nu\bar{\eta}^c_{\nu\mu}e^{-ik\cdot x_0}\,.
\end{equation}
As discussed in section \ref{subs:profile}, we now insert the gluon propagator and perform a Fourier transform, in order to obtain the result in position space. First of all we do it in the field theory limit $\alpha'k^2\rightarrow 0$. In this case we have
\begin{equation}\label{firstordertrans}
A_{\mu}^{c\,(1)}(x;\alpha'k^2\rightarrow 0)=\int \dfrac{d^4k}{(2\pi)^2}A_{\mu}^{c\,(1)}(k;\alpha'k^2\rightarrow 0)\dfrac{1}{k^2}e^{ik\cdot x}=\rho^2\bar{\eta}^c_{\nu\mu} \int \dfrac{d^4k}{(2\pi)^2}\dfrac{ik^\nu}{k^2}e^{ik\cdot(x-x_0)}\,.
\end{equation}
We remember that the scalar massless propagator in configuration space is 
\begin{equation}\label{eq:propagator}
G(x-x_0)=\int \dfrac{d^4k}{(2\pi)^2}\dfrac{1}{k^2}e^{ik\cdot(x-x_0)}=\dfrac{1}{(x-x_0)^2}\,.
\end{equation}
Deriving it with respect to $x_\nu$ we have
\begin{equation}
\partial^\nu G(x-x_0)=\int \dfrac{d^4k}{(2\pi)^2}\dfrac{ik^\nu}{k^2}e^{ik\cdot(x-x_0)}=-2\dfrac{(x-x_0)^\nu}{(x-x_0)^4}\,;
\end{equation}
going back to \eqref{firstordertrans} we can conclude that 
\begin{equation}\label{eq:profile_position_0}
A_{\mu}^{c\,(1)}(x;\alpha'k^2\rightarrow 0)=2\rho^2\bar{\eta}^c_{\mu\nu}\dfrac{(x-x_0)^\nu}{(x-x_0)^4}\,,
\end{equation}
which is exactly the leading term of the full instanton solution (with size $\rho$) in an SU(2) gauge theory \eqref{eq:singular}. It is also possible to compute $\alpha'$-correction to the profile (still in the limit $\rho\ll\sqrt{\alpha'}$): it is sufficient to perform the Fourier transform of \eqref{eq:instanton_1_momentum}, adding a the propagator in Siegel gauge. Therefore
\begin{equation}
A_{\mu}^{c\,(1)}(x)=\int \dfrac{d^4k}{(2\pi)^2}A_{\mu}^{c\,(1)}(k)\dfrac{1}{k^2}e^{ik\cdot x}= \rho^2\bar{\eta}^c_{\nu\mu} \int \dfrac{d^4k}{(2\pi)^2}\dfrac{ik^\nu}{k^2}e^{ik\cdot(x-x_0)}e^{\alpha k^2/2}\,,
\end{equation}
where $\alpha=\alpha'L_1=\alpha'\log (f'_1(0)/4)$. It can be checked that the quantity $\alpha$ is always negative, since $f'_1(0)<1$ (see below). We have
\begin{equation}
\dfrac{d}{d\alpha}A_{\mu}^{c\,(1)}(x)=\dfrac{1}{2}\partial^\nu\left[\rho^2\bar{\eta}^c_{\mu\nu}\int\dfrac{d^4k}{(2\pi)^2}e^{ik\cdot(x-x_0)}e^{\alpha k^2/2}\right]\,.
\end{equation}
The gaussian integral is easily computed, and we are left with
\begin{equation}
\dfrac{d}{d\alpha}A_{\mu}^{c\,(1)}(x)=\dfrac{1}{2}\rho^2\bar{\eta}^c_{\mu\nu}\partial^\nu\left[\dfrac{e^{(x-x_0)^2/(2\alpha'L_1)}}{(\alpha'L_1)^2}\right]\,,
\end{equation}
which integrates to
\begin{equation}
A_{\mu}^{c\,(1)}(x)=A_{\mu}^{c\,(1)}(x;\alpha'k^2\rightarrow 0)+\dfrac{1}{2}\rho^2\bar{\eta}^c_{\mu\nu}\partial^\nu\left[-2\dfrac{e^{(x-x_0)^2/(2\alpha'L_1)}}{(x-x_0)^2}\right]\,,
\end{equation}
where the integration constant is given by the result in the field theory limit \eqref{eq:profile_position_0}. Altogether the final result is
\begin{equation}
A_{\mu}^{c\,(1)}(x)=2\rho^2\bar{\eta}^c_{\mu\nu}\dfrac{(x-x_0)^\nu}{(x-x_0)^4}\left[1+e^{(x-x_0)^2/(2\alpha'L_1)}\left(1-\dfrac{(x-x_0)^2}{2\alpha'L_1}\right)\right]\,.
\end{equation}
Notice that, as expected, the correction to \eqref{eq:profile_position_0} disappears in the limit $\alpha'/(x-x_0)^2\rightarrow 0$.

\subsection*{Conformal map}
\begin{figure}[ht]
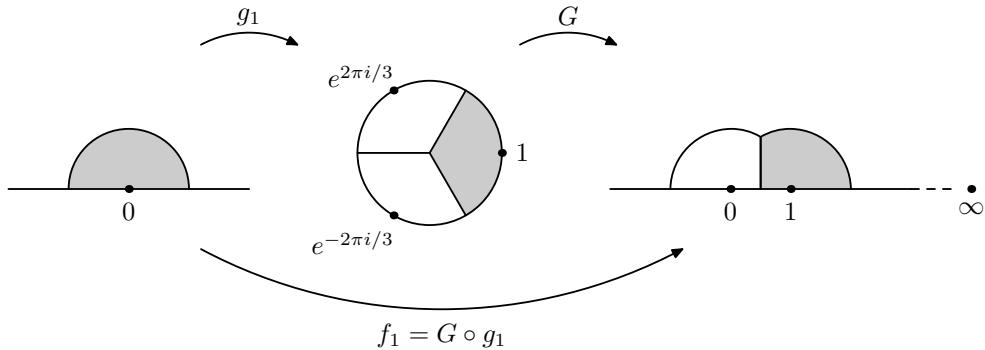

\centering
$\hbox{ \convertMPtoPDF{map.1}{1}{1} }$
\caption{Conformal map from the disk around the origin, to a disk divided in three sectors, and back to the upper half plane.}
\label{fig:Map}
\end{figure}

The function $f_1$ defined above can be obtained in two steps. First of all the unit semicircle around the origin can be mapped to a third of a disk with unit radius, with the origin mapped to the point 1. This corresponds to the map $g_1$ of figure \ref{fig:Map}, which is given by
\begin{equation}
g_1(z)=\left(\dfrac{i-z}{i+z}\right)^{2/3}\,.
\end{equation}
The second step is to map this unit disk to the upper half plane. We choose a conformal map such that the insertion points $e^{-2\pi i/3}$, 1 and $e^{2\pi i/3}$ are mapped to 0, 1 and $\infty$ respectively. Such a map is given by
\begin{equation}
G(w)=\dfrac{\left(1-e^{2\pi i/3}\right)\left(w-e^{-2\pi i/3}\right)}{\left(1-e^{-2\pi i/3}\right)\left(w-e^{2\pi i/3}\right)}\,.
\end{equation}
The map $f_1$ is then given by the composition $f_1(z)=G(g_1(z))$, and its derivative at the origin is $f'_1(0)=\frac{4}{3\sqrt{3}}<1$.
\\
It would also be possible to choose three different points $z_1$, $z_2$ and $z_3$ in \eqref{eq:prof_momentum_generic}; in that case the map $G(w)$ would have to be a different $SL(2,\mathbb{C})$ function, such that $G(e^{-2\pi i /3})=z_1$, $G(1)=z_2$ and $G(e^{2\pi i /3})=z_3$. In this way the derivative $f'(0)$ can change its value, but it turns out that the combination $f_1'(0)z_{13}/(z_{12}z_{23})$ appearing in the prefactor of \eqref{eq:prof_momentum_generic} has always norm equal to $\frac{4}{3\sqrt{3}}<1$, independently on the particular choice of $z_1$, $z_2$ and $z_3$.

\section{Derivation of the Contact Terms}\label{Pder}
In order to keep track of all the signs that arise in the graded algebra of the compositions of string vertices it is convenient to work with a shifted vector space (or a suspension). That is, we define the degree of $x$ as
$\mathrm{deg}(x)=|x|-1$, where $|x|$ is the ghost number of $x$. We denote the shifted vector space by $A[1]$ and a
shift operator $s$ as $A[1]=sA$, or
\begin{equation}
(sA)_i=A_{i-1}.
\end{equation}
where the subscript denotes the degree. With this convention, $Q$, $\xi$ and the products
\begin{equation}\label{shift1}
\begin{split}
\hat{m}_2&:=s\circ m_2\circ (s^{-1}\otimes  s^{-1})\,,\\
\hat{M}_2&:=s\circ M_2\circ (s^{-1}\otimes  s^{-1})\,,\\
\hat{M}_3&:=s\circ M_3\circ (s^{-1}\otimes  s^{-1}\otimes  s^{-1})
\end{split}
\end{equation}
all have degree one while $X$ and $\Psi_0$ have degree zero. In addition we define
\begin{eqnarray}
Q\circ \hat m_2:=Q \hat m_2\,\quad\text{and}\quad \hat m_2\circ Q:= \hat m_2\circ (Q\otimes \unit)+ \hat m_2\circ (\unit\otimes Q)
\end{eqnarray}
and similarly for $\hat M_2$, $\hat M_3$, $\hat\mu_2$ etc. Then \eqref{defM2} and \eqref{mu} become simply 
\begin{eqnarray}\label{codm}
\hat M_2=\frac{1}{3}\{X,\hat m_2\}_\circ\,=[Q,\hat \mu_2]_\circ\,,\quad\text{with}\quad  \hat \mu_2=\frac{1}{3}\{\xi,\hat m_2\}_\circ \,.
\end{eqnarray}
We are now ready to extract the contact terms in \eqref{p02} which, when expressed in terms of the maps on the shifted vector space, take the simple form
\begin{equation} \begin{split}
(\ref{p02})&=P_0\left[\hat M_2\circ Q^{-1}\circ \hat M_2-\hat M_3\right]\left(s\Psi_0\otimes s\Psi_0\otimes s\Psi_0\right)=\\
&=\sum\limits_i se^i\langle se_i,\left[\xi\circ\hat M_2\circ Q^{-1}\circ \hat M_2-\xi\circ\hat M_3\right]\left(s\Psi_0\otimes s\Psi_0\otimes s\Psi_0\right)\rangle_L\,.
\end{split} \end{equation}
Then, using \eqref{codm} we have 
\begin{equation}
\begin{split}
\langle s&e_i,\left[\xi\circ\hat M_2\circ Q^{-1}\circ \hat M_2-\xi\circ\hat M_3\right]\left(s\Psi_0\otimes s\Psi_0\otimes s\Psi_0\right)\rangle_L=\\
&=\frac{1}{2}\langle se_i,\left[\xi\circ\hat M_2\circ Q^{-1}\circ [Q,\hat \mu_2]_\circ-\xi\circ\hat M_3\right]\left(s\Psi_0\otimes s\Psi_0\otimes s\Psi_0\right)\rangle_L+\\
&\quad+\frac{1}{2}\langle se_i,\left[\xi\circ[Q,\hat \mu_2]_\circ\circ Q^{-1}\circ \hat M_2-\xi\circ\hat M_3\right]\left(s\Psi_0\otimes s\Psi_0\otimes s\Psi_0\right)\rangle_L=\\
&=\frac{1}{2}\langle se_i,\left[\xi\circ\hat M_2\circ Q^{-1}\circ Q\circ\hat \mu_2- \xi\circ\hat \mu_2\circ Q\circ Q^{-1}\circ\hat M_2\right]\left(s\Psi_0\otimes s\Psi_0\otimes s\Psi_0\right)\rangle_L+\\
&\quad+\frac{1}{2}\langle se_i,\left[X\circ\hat \mu_2\circ Q^{-1}\circ \hat M_2-2\xi\circ\hat M_3\right]\left(s\Psi_0\otimes s\Psi_0\otimes s\Psi_0\right)\rangle_L\,,
\end{split}
\end{equation}
where we used $\{Q,\xi\}_\circ=X$ and $Q e_i=0$ in the last identity. So far this calculation is identical to the calculation of the four-point scattering amplitude in \cite{Erler:2013xta}. To continue we commute $Q^{-1}$ through $Q$ using \eqref{qqm}. This is again identical to \cite{Erler:2013xta} apart form the presence of $P_0$ in \eqref{qqm}.\footnote{In \cite{Erler:2013xta} $P_0$ did not contribute due to kinematics for scattering states with finite momentum.} 
This leaves us with
\begin{equation}\label{red11}
\begin{split}
&\langle se_i,\left[-\xi\circ\hat M_2\circ P_0\circ\hat \mu_2+ \xi\circ\hat \mu_2\circ P_0\circ\hat M_2\right]\left(s\Psi_0\otimes s\Psi_0\otimes s\Psi_0\right)\rangle_L+\\
&\quad+\langle se_i,\left[X\circ\xi\circ\hat M_2\circ Q^{-1}\circ\hat m_2+X\circ\xi\circ\hat m_2\circ Q^{-1}\circ \hat M_2\right]\left(s\Psi_0\otimes s\Psi_0\otimes s\Psi_0\right)\rangle_L\,,
\end{split}
\end{equation}
where we used $\{Q,\xi\}_\circ=X$ and $Q e_i=0$ one more time and furthermore, that in the second line, the $\xi$ zero-mode has to be provided by $\hat\mu_2$. Applying \eqref{mu} once more to the second line of \eqref{red11} we are left with
\begin{equation}\label{red12}
\begin{split}
&\langle se_i,\left[-\xi\circ\hat M_2\circ P_0\circ\hat \mu_2+ \xi\circ\hat \mu_2\circ P_0\circ\hat M_2\right]\left(s\Psi_0\otimes s\Psi_0\otimes s\Psi_0\right)\rangle_L+\\
&\quad+\langle se_i,\left[X\circ\xi\circ\hat \mu_2\circ P_0\circ\hat m_2-X\circ\xi\circ\hat m_2\circ P_0\circ \hat \mu_2\right]\left(s\Psi_0\otimes s\Psi_0\otimes s\Psi_0\right)\rangle_L+\\
&\quad+\langle se_i,\left[X\circ X\circ\hat \mu_2\circ Q^{-1}\circ\hat m_2+X\circ X\circ\hat m_2\circ    Q^{-1}\circ \hat \mu_2\right]\left(s\Psi_0\otimes s\Psi_0\otimes s\Psi_0\right)\rangle_L\,.
\end{split}
\end{equation}
Before continuing, we note that the second term in the first line vanishes since in section \ref{sub:exact_marginality} we showed that $P_0M_2(\Psi_0,\Psi_0)=0$. 

To express the contribution containing $P_0$ in terms of elementary operator products we undo the shift \eqref{shift1} and use \eqref{defM2} as well as \eqref{mu}. This gives, for example,
\begin{equation}\label{red13}
\begin{split}
3\cdot&\langle se_i,\xi\circ\hat M_2\circ P_0\circ\hat \mu_2\left(s\Psi_0\otimes s\Psi_0\otimes s\Psi_0\right)\rangle_L=\\
&=\langle X\xi e_i, m_2(P_0\mu_2(\Psi_0,\Psi_0),\Psi_0) \rangle_L  + \langle\xi e_i,m_2(XP_0\mu_2(\Psi_0,\Psi_0),\Psi_0)  \rangle_L+\\
&\quad+\langle\xi e_i,m_2(P_0\mu_2(\Psi_0,\Psi_0),X\Psi_0)  \rangle_L
    +\langle X\xi e_i, m_2(\Psi_0,P_0\mu_2(\Psi_0,\Psi_0)) \rangle_L +\\
    &\quad+ \langle\xi e_i,m_2(X\Psi_0,P_0\mu_2(\Psi_0,\Psi_0))  \rangle_L+\langle\xi e_i,m_2(\Psi_0,XP_0\mu_2(\Psi_0,\Psi_0))  \rangle_L\,,
\end{split}
\end{equation}
where we used the fact that $\xi$ and $X$ are both BPZ even. 
The remaining terms in the second line of \eqref{red12} give in turn 
\begin{equation}\label{PA11}
\begin{split}
    &-\langle X \xi e_i, m_2(P_0\mu_2(\Psi_0,\Psi_0),\Psi_0)+ m_2(\Psi_0, P_0\mu_2(\Psi_0,\Psi_0))\rangle_L+\\
    &-\langle X \xi e_i, \mu_2(P_0m_2(\Psi_0,\Psi_0),\Psi_0)- \mu_2(\Psi_0, P_0m_2(\Psi_0,\Psi_0))\rangle_L\,.
\end{split}
\end{equation}
Making use of the cyclic properties of $m_2$ and $\mu_2$ (e.g. \cite{Moeller:2010mh}) 
\begin{equation}\label{cyclic}
\begin{split}
&\langle a,m_2(b,c)\rangle=(-1)^{|a|(|b|+|c|)} \langle b,m_2(c,a)\rangle\,,\\
&\langle a,\mu_2(b,c)\rangle=(-1)^{|a|+|b|+|a|(|b|+|c|)} \langle b,\mu_2(c,a)\rangle
\end{split}
\end{equation}
we can recast \eqref{red13} and \eqref{PA11} into 
\begin{equation} \begin{split}
    &-\frac{1}{3}\langle P_0\mu_2(\Psi_0,\Psi_0), 4m_2(\Psi_0,\xi Xe_i)-4m_2(\xi Xe_i,\Psi_0) +  m_2(X\Psi_0,\xi e_i)-m_2(\xi e_i,X\Psi_0) \rangle_L+\\
    &-\frac{1}{3}\langle XP_0\mu_2(\Psi_0,\Psi_0), m_2(\Psi_0,\xi e_i) - m_2(\xi e_i,\Psi_0)\rangle_L+\\
    &-\langle P_0m_2(\Psi_0,\Psi_0), \mu_2(\Psi_0,\xi Xe_i) - \mu_2(\xi Xe_i,\Psi_0)\rangle_L\,,
\end{split} \end{equation}
where we have furthermore used that $X\xi A=\xi XA$ for $A=V,e_i$. Finally, we can use the definition of the product $\mu_2$ in the last line, and arrive at the result that we quote in \eqref{P02b}.

This leaves us with the terms containing the propagator $Q^{-1}$ \eqref{red12}. Here, the $\xi$ zero-mode has to be provided by $\mu_2$. Thus, the last line in \eqref{red12} can be written in the small Hilbert space as 
 \begin{eqnarray}\label{redws_app}
-2\langle X\circ X\,e_i, m_2( Q^{-1}m_2(\Psi_0,\Psi_0),\Psi_0) + m_2(\Psi_0,Q^{-1} m_2(\Psi_0, \Psi_0)\rangle\,.
\end{eqnarray}

\section{Anomalous Contributions due to Non-Primary Fields}\label{app:anomaly}
The explicit calculation of the product $m_2$ and of the BPZ inner product requires a set of conformal transformation that map each vertex operator to the upper half plane. In this way the product $m_2$ can be expressed in terms of the operator product expansion of operators in CFT, and the BPZ inner product is equivalent to a correlation function on the upper half plane. If all the vertex operators are primaries of conformal dimension $0$ this does not pose any problem. In this paper, however, we are dealing with some non-primaries operators, hence we should consider anomalous contributions due to these conformal transformations. Let us consider an operator $W$ of scaling dimension $h=0$, but with anomalous OPE with the energy-momentum tensor given by
\begin{equation}
T(z)W(0)=\dfrac{\alpha}{z^3}+\dfrac{\partial W(0)}{z}+\dots
\end{equation}
Considering now an infinitesimal transformation $z\rightarrow z+\epsilon(z)$, the operator $W$ transforms according to
\begin{equation}\label{eq:deltaW}
\delta_\epsilon W(w)=\dfrac{1}{2\pi i}\oint dz[T(z)\epsilon(z),W(w)] =\alpha\partial^2\epsilon(w)W(w)+\epsilon(w)\partial W(w)\,.
\end{equation}
The last term in \eqref{eq:deltaW} would be present even if the operator $W$ was primary, while the first term $\epsilon''(z)W(z)$ is an anomalous contribution.

In the bulk of the paper we have encountered two non-primary operators, namely $\frac{1}{4}:\xi\eta:e^{\phi}\widetilde{\mathbb{V}}_{1/2}$ and $\partial(e^{\phi}\widetilde{\mathbb{V}}_{1/2})$ (see appendix \ref{app:OPE}). The first one appearns in $\xi X e_i$ and gives anomalous contributions to \eqref{P02b}. This anomalous contribution will thus be equal, using \eqref{eq:Txieta}, to
\begin{equation}\label{eq:anomaly1}
-\dfrac{6}{3}\dfrac{1}{4}(\partial^2\epsilon_3(0)-\partial^2\epsilon_4(0))\langle\xi P_0m_2(\Psi_0,\Psi_0), m_2(\Psi_0,e^{\phi}\widetilde{\mathbb{V}}_{1/2})-m_2(e^{\phi}\widetilde{\mathbb{V}}_{1/2},\Psi_0)\rangle_L\,,
\end{equation}
where $\epsilon_{3,4}$ represent the infinitesimal part of the two conformal transformation that have to be done in order map the BPZ product to a correlation function on the upper half plane. On the other hand, $\partial(e^{\phi}\widetilde{\mathbb{V}}_{1/2})$ gives anomalous contributions to \eqref{eq:propagator_fin}. Using \eqref{eq:Tderiv} we find that the anomaly is given by
\begin{equation}
\dfrac{1}{4}(2\partial^2\epsilon_3(0)-2\partial^2\epsilon_4(0))\langle P_0m_2(\Psi_0,\Psi_0), m_2(\Psi_0,e^{\phi}\widetilde{\mathbb{V}}_{1/2})-m_2(e^{\phi}\widetilde{\mathbb{V}}_{1/2},\Psi_0)\rangle\,,
\end{equation}
which exactly cancels the other anomalous contribution \eqref{eq:anomaly1}.

\cleardoublepage
\addcontentsline{toc}{section}{References}
\bibliography{bibliography}
\bibliographystyle{JHEP}

\end{document}